\documentclass[12pt]{article}

\usepackage[a4paper,text={16.8cm,22.4cm}]{geometry}
\usepackage{amsmath,amsfonts,slashed,amssymb,tikz,bm,psfrag,graphicx,color,dsfont,soul, multicol,multirow,ulem}

\RequirePackage[sort&compress,square,comma,numbers]{natbib}
\allowdisplaybreaks
\usepackage{amsthm,xcolor,float}

\usepackage{esint}
\usepackage[unicode=true]{hyperref}
\hypersetup{
     colorlinks=true,       		
     linkcolor=blue,          	
     citecolor=red,            
     urlcolor=magenta,           	
 }

\usepackage{mathrsfs,mathtools,mathdots,dsfont}
\usepackage{listings}
\usepackage{subfigure}
\usepackage{makecell}
\usepackage{multirow}

\newtheorem*{theorem*}{Theorem}

\newtheorem*{corollary*}{Corollary}

\newtheorem*{lemma*}{Lemma}

\newtheorem*{proposition*}{Example*}

\newtheorem*{conjecture*}{Conjecture}
\theoremstyle{definition}

\newtheorem*{definition*}{Definition}
\theoremstyle{remark}

\newtheorem*{remark*}{Remark}

\newcommand{\e}{\mathrm{e}}

\renewcommand{\d}{\mathrm{d}}

\newcommand{\ket}[1]{\left|#1\right\rangle}
\newcommand{\bra}[1]{\left\langle#1\right|}
\newcommand{\Eq}[1]{{\rm Eq.}~\eqref{#1}}

\begin{document}

\begin{titlepage}

\vspace{0.1cm}
\begin{center}
\Large\bf
Next-to-Leading Order QCD Corrections to $\Lambda_b \to p $ Form Factors from  Light-Cone Sum Rules  
\end{center}

\vspace{0.5cm}
\begin{center}
{\bf Jiang-Lin Zhou$^{a}$ and Yong-Kang Huang$^a$} \\
\vspace{0.7cm}
{\sl  ${}^a$\, School of Physics, Nankai University, Tianjin 300071, People's Republic of China\\
}
\end{center}

\date{\today}

\vspace{0.2cm}

\begin{abstract}
In this study, we compute the radiative corrections to the $\Lambda_b \to p$ transition form factors at next-to-leading logarithmic accuracy, employing the framework of QCD light-cone sum rules with the light-cone distribution amplitudes of the $\Lambda_b$ baryon.
 The factorization formulae of the vacuum-to-$\Lambda_b$ correlation function, constructed from the interpolating current for the proton, are derived at leading power in $m_p / m_{\Lambda_b}$, using the method of regions.
With our specific choice of interpolating current, only the twist-4 distribution amplitude of the $\Lambda_b$ baryon contributes to the form factors. 
Numerically, we find that the next-to-leading order QCD perturbative corrections  reduce the tree-level form factors  to approximately 65$\%$ of their original value, with the next-to-leading-order jet function providing the dominant contribution.
In the large-energy limit ($E_p \to \infty$), the form factors exhibit a clear $1/E_p^3$ scaling, consistent with the expected power-counting behavior. 
By applying the $z$-series parameterization to perform a combined fit of the form factors from our results and available lattice QCD simulations, we further investigate the decay rate of $\Lambda_b \to p \ell^- \bar{\nu}_{\ell}$ and extract the CKM matrix element $|V_{ub}| = (3.33\pm 0.43 ) \times 10^{-3}$.
	

\end{abstract}
\vfil

\end{titlepage}

\section{ Introduction}

The decays $\Lambda_b \to p \ell^- \bar{\nu}_{\ell}$ and $\Lambda_b \to \Lambda_c \ell^- \bar{\nu}_{\ell}$ provide important channels for determining the Cabibbo-Kobayashi-Maskawa (CKM) matrix element $|V_{ub}|$ and $|V_{cb}|$ \cite{detmold_mathrmensuremathlambda_bensuremathrightarrowpensuremathellensuremath-overlineensuremathnu_ensuremathell_2015}. Using the decay-width measurements from LHCb \cite{aaij_determination_2015} together with lattice QCD results to extract $|V_{ub}|/|V_{cb}|$, one obtains $|V_{ub}|=(3.27 \pm 0.15 \pm 0.16\pm 0.06 ) \times 10^{-3}$ \cite{aaij_determination_2015}. This value lies about $2\sigma$ below the exclusive $B$-decay determination, $ (3.70 \pm 0.22 ) \times 10^{-3}$ \cite{ParticleDataGroup:2024cfk}.
In addition, the observed discrepancy between exclusive and inclusive determinations of $|V_{ub}|$ in $B$ decays is a long-standing tension \cite{ParticleDataGroup:2024cfk}. Resolving these discrepancies requires both further experimental input and more precise theoretical calculations of the relevant heavy-to-light transition form factors.
In this context, achieving accurate theoretical predictions for the $\Lambda_b \to p$ form factors is of particular importance. They are essential both for extracting $|V_{ub}|$ \cite{detmold_mathrmensuremathlambda_bensuremathrightarrowpensuremathellensuremath-overlineensuremathnu_ensuremathell_2015,faustov_semileptonic_2016} and for reliably evaluating various non-leptonic $\Lambda_b$ decay channels, such as $\Lambda_b \to p K^-$ \cite{PhysRevD.80.034011,hsiao_roles_2016,zhu_decay_2016}.


The precise calculation of heavy-to-light transition form factors is fundamentally challenged by the need to account for nonperturbative QCD dynamics within the bound-state hadrons.
 To address this, several theoretical frameworks have been developed. These include effective theories such as Heavy Quark Effective Theory (HQET) \cite{isgur_heavy-baryon_1991,mannel_baryons_1991,hussain_general_1992} and Soft-Collinear Effective Theory (SCET) \cite{PhysRevD.63.114020,PhysRevD.65.054022,beneke_soft-collinear_2002,PhysRevD.85.014035,lu_scet_2025}, as well as factorization approaches like the perturbative QCD  approach (pQCD) \cite{PhysRevD.80.034011,PhysRevD.75.054017,PhysRevD.63.074009,keum_fat_2001,han_lambda_2022}. 
 While originally advanced in the context of heavy meson decays, these methods have inspired significant applications to heavy baryon systems, though their implementation often differs from the mesonic case.
For $\Lambda_b \to p$ form factors, SCET analyses indicate that the heavy-to-light form factors are factorizable at leading power, yet their numerical magnitude can be suppressed relative to power-suppressed soft contributions \cite{wang_factorization_2012}. 
Complementary pQCD evaluations of the $\Lambda_b \to p$ transition reveal that the dominant contributions arise from the twist-4 $\Lambda_b$ baryon distribution amplitude and the twist-4 and twist-5 light-cone distribution amplitudes of the proton \cite{han_lambda_2022}.

Nonperturbative methods also play an essential role in studying heavy-to-light form factors. Notable examples include lattice QCD, QCD sum rules, light-cone sum rules (LCSR), and the light-front quark model. 
Lattice QCD, grounded in the first principles of QCD, is widely regarded as a reliable tool and has been employed to study $\Lambda_b \to p \ell^- \bar{\nu}_{\ell}$ and $\Lambda_b \to \Lambda_c \ell^- \bar{\nu}_{\ell}$ decays \cite{detmold_mathrmensuremathlambda_bensuremathrightarrowpensuremathellensuremath-overlineensuremathnu_ensuremathell_2015,PhysRevD.93.054003}. 
Its calculations are typically most robust in the low-recoil (large $q^2$) region. To obtain precise predictions across the full kinematic range, phenomenological extrapolations from lattice data are often necessary. 
Here, LCSR proves to be a valuable complementary approach, as it can provide relatively precise results in the large-recoil (low $q^2$) region.

LCSR is a powerful framework for studying hard exclusive processes. Its core methodology is to expand the products of currents near the light-cone .
 For heavy-to-light transitions, it is to  perform an operator product expansion (OPE) of a correlation function constructed from interpolating currents near the light-cone, and subsequently factorize it into perturbatively calculable kernels and non-perturbative light-cone distribution amplitudes (LCDAs) of the relevant hadrons \cite{PhysRevD.63.114020, beneke_symmetry-breaking_2001, bonciani_two-loop_2008, wang_perturbative_2016, PhysRevD.101.074035}. This framework facilitates a model-independent treatment of both hard and soft contributions to form factors. The flexibility in choosing different interpolating currents and distribution amplitudes provides a systematic way to study hadronic matrix elements.
The choices of  the LCDAs of the light baryons or the heavy one leads to two primary implementations, namely light-hadron LCSR \cite{khodjamirian_form_2011}  and heavy-hadron LCSR \cite{wang_ensuremathlambda_bensuremathrightarrowp_2009,wang_perturbative_2016,huang_lambda_2023,miao_b_2022}. The light-hadron LCSR  employs the LCDAs of the final-state light baryon \cite{khodjamirian_form_2011}, and the heavy-hadron LCSR uses the LCDAs of the initial heavy baryon \cite{wang_ensuremathlambda_bensuremathrightarrowp_2009, wang_perturbative_2016, huang_lambda_2023, miao_b_2022}. For the $\Lambda_b \to p$ transition, many  studies using LCSR have been conducted. However, they have largely remained at the leading-order (LO) accuracy \cite{wang_ensuremathlambda_bensuremathrightarrowp_2009, khodjamirian_form_2011, huang_lambda_2023}. Studies of $B$-meson decays \cite{wang_subleading_2017, wang_qcd_2015} and the $\Lambda_b \to \Lambda \ell^+ \ell^-$ process \cite{wang_perturbative_2016} have demonstrated that next-to-leading order (NLO) QCD corrections can induce reductions of approximately 30$\%$  and 50$\%$ respectively, highlighting their significant impact on precision calculations.

Therefore, in this work, we calculate the QCD radiative corrections to the $\Lambda_b \to p$ form factors at next-to-leading logarithmic (NLL) accuracy within the heavy-hadron LCSR framework.
We employ the  leading-power (LP) current operator \cite{PhysRevD.73.094019, han_lambda_2022} to construct the vacuum-to-$\Lambda_b$ correlation function. 
As we will demonstrate,  only the twist-4 $\Lambda_b$ LCDA contributes to the factorization formula at leading power. To handle its unknown one-loop renormalization group (RG) evolution, we adapt the technique developed in \cite{wang_perturbative_2016}. After resumming large logarithms via the standard RG approach, we obtain the $\Lambda_b \to p$ form factors at NLL accuracy. We then perform a detailed numerical analysis, evaluate key phenomenological observables for the decay $\Lambda_b \to p \ell^- \bar{\nu}_{\ell}$, and extract $|V_{ub}|$.

This paper is organized as follows. 
In section \ref{sect2}, we introduce the definitions of the $\Lambda_b \to p$ form factors and construct the vacuum-to-$\Lambda_b$ correlation function. We derive the tree-level LCSR and analyze the power counting of the form factors. Section \ref{sect3} details the calculation of the one-loop corrections. Using the method of regions, we extract the hard and jet functions and demonstrate the cancellation of the factorization-scale dependence. By comparing with the $\Lambda_b \to \Lambda$ form factors, we analyze the implications of flavor symmetry. The final NLL expressions for the $\Lambda_b \to p$ form factors are presented. In section \ref{sect4}, we perform the numerical analysis. We discuss the input parameters, model dependencies, and the observed ~35$\%$ reduction from NLL corrections. The behavior of the form factors with the proton energy $E_p$ is shown, and the dominant role of the jet function is observed. Using the $z$-series expansion, we perform a combined fit to our LCSR results and the lattice QCD data \cite{detmold_mathrmensuremathlambda_bensuremathrightarrowpensuremathellensuremath-overlineensuremathnu_ensuremathell_2015} to describe the form factors across the full kinematic range. These fitted results are used for phenomenological applications, including the prediction of $|V_{ub}|$. A concluding discussion is given in section \ref{sect5}.

\section{Tree-level LCSR of the $\Lambda_b \to p$ form factors}\label{sect2}
\subsection{Helicity-based $\Lambda_b \to p$ form factors }
\begin{figure}[htbp]
					\centering
					\includegraphics[scale=1]{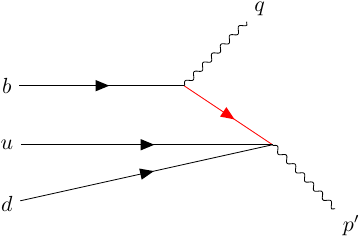}
                    \caption{Diagrammatical representation of the correlation function $\Pi_{\mu,a}(n\cdot p',\bar{n} \cdot p')$ at tree level, where the red internal line indicates the hard-collinear propagator of the up quark, $q$ represents the transfer momentum caused by the weak transition vertex, and $p'$ represents the proton caused by the local interpolating current operator. }
		\end{figure}
In this work, we define $\Lambda_b \to p $ form factors in the helicity basis. We collect the definitions as follows  \cite{detmold_mathrmensuremathlambda_bensuremathrightarrowpensuremathellensuremath-overlineensuremathnu_ensuremathell_2015} ,
\begin{align}
    \bra{{p} (p',s')} \bar{u} \gamma_{\mu} b \ket{\Lambda_b (P,s)} = & {\bar{u}_p} (p',s') \left [ f^0_{\Lambda_b \to p } (q^2 ) \frac{m_{\Lambda_b } - m_p  }{q^2 } q_{\mu} \right .
\nonumber \\
    & + f^+_{\Lambda_b \to p } (q^2 ) \frac{m_{\Lambda_b}+m_p }{s_+} \left( P_\mu + p'_\mu - \frac{m^2_{\Lambda_b} -m^2_p}{q^2} q_{\mu}  \right ) \notag\\
    & \left . + f^T_{\Lambda_b \to p } (q^2) \left ( \gamma_{\mu} - \frac{2 m_p }{s_+}P_{\mu} -\frac{2 m_{\Lambda_b} }{s_+}p'_{\mu}  \right )  \right ]   u_{\Lambda_b} (P,s),
\label{hadV} \\
\nonumber \\
    \bra{p(p' ,s')} \bar{u} \gamma_{\mu}  \gamma_5 b \ket{\Lambda_b (P,s )} = & - {\bar{u}_p } (p',s')  \gamma_5 \left [  g^0_{\Lambda_b \to p } (q^2 ) \frac{m_{\Lambda_b } + m_p  }{q^2 } q_{\mu} \right .
\nonumber \\
    & + g^+_{\Lambda_b \to p } (q^2) \frac{m_{\Lambda_b}-m_p }{s_-}\left( P_{\mu} + p'_\mu - \frac{m^2_{\Lambda_b} -m^2_p}{q^2} q_{\mu} \right )
\nonumber \\
    & \left . + g^T_{\Lambda_b \to p } (q^2)\left ( \gamma_{\mu} + \frac{2 m_p }{s_-}P_{\mu} -\frac{2 m_{\Lambda_b} }{s_-}p'_{\mu} \right ) \right ]  u_{\Lambda_b} (P,s) , 
 \label{hadA}
\end{align}
where $p$ represents proton , $m_{\Lambda_b} (s ) $ is the mass (spin) of the $\Lambda_b - $baryon ,  $m_{p} (s' ) $ is the mass (spin) of proton, and we designate 
\begin{eqnarray}
 q_{\mu} = P_{\mu} - p'_{\mu} \,,\,   s_{\pm} = (m_{\Lambda_b} \pm m_p )^2 - q^2 .
\end{eqnarray}  
We work in the rest frame of the $\Lambda_b$-baryon with the velocity vector $v_{\mu} = P_{\mu}/ m_{\Lambda_b} $ and define a light-cone vector $\bar{n}_{\mu}$ parallel to the four-momentum $p'$ of proton in the massless limit. 
The other light-cone vector can be introduced as $n_{\mu} = 2 v_{\mu} - \bar{n}_{\mu} $  with $ n^2 = 0 \, , \bar n^2 = 0 \, , n \cdot \bar{n} =2 $. 

For simplicity, we designate $p'$ as 
\begin{eqnarray}
    p'_{\mu} = \frac{n \cdot p'}{2}\bar{n}_{\mu}+\frac{\bar{n} \cdot p'}{2}{n}_{\mu},
\end{eqnarray}
and at  large hadronic recoil, we assume $\bar{n} \cdot p' $ is so smaller than $n \cdot p' $ that  we  have 
\begin{eqnarray}
    n \cdot p' \simeq \frac{m^2_{\Lambda_b} + m^2_p -q^2 }{m_{\Lambda_b}}  = 2 E_p \sim \mathcal{O}(m_{\Lambda_b}),
\end{eqnarray} 
and $\bar{n} \cdot p' \sim \mathcal{O }(\frac{m_p^2}{m_{\Lambda_b}})  \sim \lambda m_{\Lambda_b}$, where $\lambda \sim \frac{\Lambda_{\rm QCD}}{ m_{\Lambda_b} }$ .
We will conduct our research at leading power in $\lambda$.


\subsection{Interpolating current operators and correlation function}
Following the standard strategy \cite{wang_perturbative_2016}, we construct a vacuum-$\Lambda_b$-baryon correlation function, 
\begin{eqnarray}\label{correlation}
    \Pi_{\mu,a} (P,p') = i \int d^4 x e^{i  p' \cdot x } \bra{0} T \{ j_p (x) , j_{\mu,a} (0) \} \ket{\Lambda_b (P)},
\end{eqnarray}
 where the local current operator $j_p $   interpolates the proton and the current $j_{\mu,a}$ denotes the weak transition current, with $a = V, A$ labeling the vector and axial-vector currents
 \begin{align}
    j_{\mu,V} = & \, \bar{u} \gamma_{\mu } b \,, & j_{\mu,A} = & \, \bar{u} \gamma_{\mu }  \gamma_5 b.
 \end{align}

 Similar with the $\Lambda$-baryon \cite{khodjamirian_form_2011,wang_perturbative_2016}, the general  structure of the proton interpolating current operator is 
 \begin{eqnarray}
    j_p= \varepsilon_{i,j,k} [u_i^T C \Gamma u_j] \,    \tilde{\Gamma} d_k \, , 
 \end{eqnarray}
 where $C$ is the charge conjugation matrix and the sum runs over the color indices $i,j,k$.

 Implementing the isospin constraint of the light diquark $[ud] $ system of the proton, there exist three local operators that do not involve derivatives \cite{ioffe_choice_1983}, namely LP current operator $j_1$ \cite{chernyak_nucleon_1984,gockeler_generalized_2005,huang_lambda_2023}, the tensor current operator $j_2$ \cite{huang_lambda_2023} and the Ioffe current operator $j_3$ \cite{ioffe_calculation_1981,braun_higher_2000},
 \begin{align} \label{Jp}
     j_{1} = & \varepsilon_{i,j,k} \left [ u_i^T C \slashed n u_j \right ] \gamma_5 \slashed n d_k   \,, &   j_{2} = & \varepsilon_{i,j,k}  \left [ u_i^T C \gamma_{\mu} u_j \right ]   \gamma_5 \gamma^{\mu}  d_k  \,, & j_{3}= & \varepsilon_{i,j,k} \left [ u_i^T C \sigma_{\mu \nu } u_j \right ]   \gamma_5 \sigma^{\mu \nu }  d_k \, ,
 \end{align}
 where the latter two current operators are widely used in calculations of dynamical
characteristics of the nucleon in the QCD sum rule approach.

Because the large component of the momentum of the proton $p'_{\mu}$ is along $\bar{n}_{\mu}$ and the large-momentum component of the quarks constituting the proton should also align with the $\bar{n}$-direction.   
So we can introduce two projection operators 
\begin{align}
	P_+ = & \frac{\slashed{\bar n} \slashed n}{4}\,, &  P_- = & \frac{  \slashed n \slashed{\bar n} }{4} \,, &  P_+ +P_- =1 \, ,
\end{align}
which project onto the "plus" and "minus" components of the spinor, like the large and small components of the (hard)-collinear quark fields in SCET \cite{becher_generalization_2015}.
  We can find that the two current operator $j_2$ and $j_3 $ are power suppressed compared with $j_1$ \cite{huang_lambda_2023} . 

 Although there is no general recipe to discriminate various choices for the interpolating field of baryon.
 a practical criterion is that the coupling between the interpolating current operator and the given state should be strong enough that one can minimize the contamination generated by its coupling to the unwanted hadronic states and  poorly known contributions of higher dimension(twist) operators \cite{PhysRevD.73.094019}. 
 Compared with the others listed in \Eq{Jp}, $j_1$ is not power suppressed. 
 So we will only consider $j_1$ for construction of the correlation function to conduct the conduct the analysis of the next-to-leading-order QCD corrections at leading power.

By inserting a complete set of hadronic states, one can directly derive the hadronic spectral representation of the correlation function defined with the vector and axial-vector currents at leading power
\begin{align}
    \Pi_{\mu ,V }(P,q) = & \frac{f_{N}(\mu) \, n \cdot p' }{m_p^2 / n \cdot p' - \bar{n} \cdot p' - i0  } \left [ 2 f^T_{\Lambda_b \to p } (q^2) \gamma_{\perp \mu}  + {m_{\Lambda_b} \over m_{\Lambda_b} - n\cdot p'} \left [ f^0_{\Lambda_b \to p } (q^2) - f^+_{\Lambda_b \to p } (q^2) \right ] n_{\mu}  \right .
\nonumber \\
    & + \left [ f^0_{\Lambda_b \to p } (q^2)+f^+_{\Lambda_b \to p } (q^2) \right ] {\bar{n}}_{\mu} \Big ]  \, {\slashed n \over 2} u_{\Lambda_b} (P) + \int_{\omega_s}^{\infty} d \omega \,  \frac{1}{\omega- \bar{n} \cdot p' - i0  }
\nonumber \\
    & \times \left[ {\rho}^h_{V, \perp } (\omega, n \cdot p' ) \gamma_{\perp \mu} + {\rho}^h_{V, n } (\omega, n \cdot p' ) n_{\mu} + {\rho}^h_{V, \bar{n} } (\omega, n \cdot p' )  \bar{n}_{\mu} \right] \, \slashed n u_{\Lambda_b} (P ) \, , \label{hadpV}
\\
    \Pi_{\mu ,A }(P,q) = & \frac{f_{N}(\mu) \, n \cdot p' }{m_p^2 / n \cdot p' - \bar{n} \cdot p' - i0  } \left [ 2 g^T_{\Lambda_b \to p } (q^2) \gamma_{\perp \mu}  - {m_{\Lambda_b} \over m_{\Lambda_b} - n\cdot p'} \left [ g^0_{\Lambda_b \to p } (q^2) - g^+_{\Lambda_b \to p } (q^2) \right ] n_{\mu}  \right .
\nonumber \\
    & - \left [ g^0_{\Lambda_b \to p } (q^2) + g^+_{\Lambda_b \to p } (q^2) \right ] {\bar{n}}_{\mu} \Big ] \gamma_5  \, {\slashed n \over 2} u_{\Lambda_b} (P) + \int_{\omega_s}^{\infty} d \omega \,  \frac{1}{\omega- \bar{n} \cdot p' - i0  }
\nonumber \\
    & \times \left[ {\rho}^h_{A, \perp } (\omega, n \cdot p' ) \gamma_{\perp \mu} - {\rho}^h_{A, n } (\omega, n \cdot p' ) n_{\mu} - {\rho}^h_{A, \bar{n} } (\omega, n \cdot p' )  \bar{n}_{\mu} \right] \, \gamma_5 \slashed n u_{\Lambda_b} (P ) \, , \label{hadpA}
\end{align}
where the proton decay constant $f_N(\mu)$, which is dependent on the scale of renormalization, is defined by the hadronic matrix element of $j_1$ 
\begin{eqnarray}\label{Interpolating}
    \bra{0} j_{1 }(0) \ket{p(p')} = f_{N}(\mu) \, n \cdot p' \, \slashed n  u_p(p') \, .
 \end{eqnarray}
The RGE of the decay constant is given by
\begin{eqnarray}\label{RGfn}
	\frac{d }{d \ln \mu} \ln f_N(\mu) = - \sum_{k=0} \left(\frac{\alpha_s (\mu)}{4 \pi}\right)^{k+1} \gamma_{N}^{(k)} ,
\end{eqnarray}
where the anomalous dimension $\gamma_N^{(k)}$ have been calculated at two-loop level in general $\overline{\rm MS}$ scheme and KM scheme \cite{PhysRevD.111.094517,huang2026}.
\subsection{Tree-level LCSR}

Now for the correlation function $\Pi_{\mu,a } (P,q)$ at space-like interpolating momentum with $\bar{n} \cdot p' \sim \lambda \, m_{\Lambda_b}$ and $n \cdot p' \sim  m_{\Lambda_b}$, $x^2 \to 0$. So light-cone operator-product-expansion (OPE) is applicable. The correction function $\Pi_{\mu,a}$ can be factorized into
\begin{eqnarray}\label{treecf}
    \Pi_{\mu, a} (P,q) = \int_0^\infty d \omega_1 d \omega_2 \Big \{ T_{\perp }(\omega_i) \left \langle  O_{4\perp, a, \mu}(\omega_i) \right \rangle + \left [ T_{n }(\omega_i )\, n_{\mu} + T_{\bar{n} }(\omega_i )\, {\bar{n}}_{\mu} \right ]  \left \langle  O_{4 \parallel,a}(\omega_i) \right \rangle \Big \} \, , \hspace{0.4cm}
\end{eqnarray}
 where $T_{\perp }\,,\, T_{n } \,,\, T_{\bar{n} }$ are the perturbatively calculable short-distance coefficients, independent of soft hadronic states. Therefore, we are able to compute the correlation function at the quark level to extract the short-distance coefficients.
The non-perturbative hadronic matrix elements are defined by the non-local operators in HQET,
\begin{align}
	\left \langle  O_{4\perp, a, \mu} (\omega_1, \omega_2) \right \rangle = &\int {d t_1 \over 2 \pi} {d t_2 \over 2 \pi} \e^{i (\omega_1 t_1 +\omega_2 t_2 )} \varepsilon_{ijk} \notag\\
	& \left \langle 0 \left |  \Big [ u_i^{T}(t_1 \bar{n}) \, [0,t_1 \bar{n}] \, C   \gamma_5 \slashed n d_j (t_2 \bar{n})  [0,t_2 \bar{n}] \Big ] \, \gamma_{\perp\mu} \left(1 \,,\gamma_5\right) \slashed n \, h_{v,k}(0) \right | \Lambda_b(v) \right \rangle \, ,
\nonumber \\
	\left \langle  O_{4\parallel,a} (\omega_1 , \omega_2) \right \rangle = & \int {d t_1 \over 2 \pi} {d t_2 \over 2 \pi} \e^{i(\omega_1' t_1 +\omega_2' t_2 )} \varepsilon_{ijk} \notag\\
	& \left \langle 0 \left | \Big [ u_i^{T}(t_1 \bar{n})] [0,t_1 \bar{n}] C \gamma_5 \not{n}  d_j(t_2 \bar{n}) [0,t_2 \bar{n}] \Big ]  \, \left(1\, , - \gamma_5\right)  \slashed n h_{v,k}(0) \right | \Lambda_b(v) \right \rangle \, ,
\end{align}  
where the light-cone Wilson line 
 \begin{eqnarray}
    [0, t \bar{n}] = {\rm P} \left\{{\rm Exp}\left[-i \, g_s \, t \int_{0}^{1} du \bar{n} \cdot A(u \, t \, \bar{n}) \right] \right\}
 \end{eqnarray}
 is introduced to maintain gauge invariance in the collinear direction.

 With the help of the most-general light-cone hadronic matrix element of $\Lambda_b$-baryon in coordinate space \cite{bell_light-cone_2013} , we find the matrix elements of the operators $O_{4\perp,a,\mu}$ and $O_{4\parallel,a}$ can define the same twist-4 LCDA of $\Lambda_b-$baryon $\phi_4(\omega_1' , \omega_2' ) $, 
 \begin{align}
	\left \langle  O_{4\perp, a, \mu} (\omega_1,\omega_2) \right \rangle = &f^{(2)}_{\Lambda_b}(\mu) \phi_4(\omega_1, \omega_2) \, \, [ \gamma_{\perp\mu} \, \left(1\, , \gamma_5\right) \, \slashed n u_{\Lambda_b} (p') ] \, ,
\nonumber \\
	\left \langle  O_{4\parallel,a}  (\omega_1,\omega_2)  \right \rangle = & f^{(2)}_{\Lambda_b}(\mu) \phi_4(\omega_1 , \omega_2)  \, \, [ \left(1\, , - \gamma_5\right) \slashed n u_{\Lambda_b} (p') ].
\end{align}
 Through the calculation of the tree-level correlation function with quark external states, the leading-order hard kernels in \Eq{treecf} can be directly extracted as
\begin{align}
	T^{(0)}_{\perp } = & \, T^{(0)}_{\bar{n} } = \frac{1}{ \omega_1 + \omega_2 - \bar{n} \cdot p'}  \,, &  T^{(0)}_{n } = & \,  0 \, .
\end{align}

It is now straightforward to derive the tree-level factorization formulae at leading power in $\lambda $, 
\begin{eqnarray}
     \Pi_{\mu,V(A)} (P,p') = f^{(2)}_{\Lambda_b} (\mu) \int_{0}^{\infty} d \omega_1 d \omega_2 \frac{ \phi_4(\omega_1, \omega_2, \mu) }{ \omega_1 + \omega_2 - \bar{n} \cdot p' - i 0 } \left [ (1,- \gamma_5)  \left(\gamma_{\perp \mu} + \bar{n}_{\mu} \right) \slashed n u_{\Lambda_b} (p') \right ] \, .
\end{eqnarray}

 Employing the parton-hadronic duality approximation for the dispersion integrals in the hadronic representations and performing the continuum subtraction
 as well as the Borel transformation, we obtain the tree-level LCSR 
 \begin{eqnarray}
    F^i_{\Lambda_b \to p}(q^2 ) = \frac{ f^{(2)}_{\Lambda_b}(\mu)  }{ f_N(\mu)  n \cdot p' } {\rm exp } \left[\frac{m_p^2}{n \cdot p' \omega_M}\right]  \int_{0}^{\omega_s}  d \omega \, e^{-{\omega \over \omega_M}} \, \tilde{\psi}_4 (\omega)  + \mathcal{O}( \alpha_s) \, ,
 \end{eqnarray}
where $ F^i_{\Lambda_b \to p}(q^2 ) $ represents any of the six $\Lambda_b \to p $ form factors defined in \Eq{hadV}, \Eq{hadA} and $\omega_1 = u \, \omega$,  $\omega_2 = \bar u \, \omega$,
\begin{align}
    \tilde \psi_4 (\omega) = & \omega \int_0^1 d u \, \phi_4 (u \, \omega \, , \bar u \, \omega) \, , & \bar u = & 1 - u \, .
\end{align}

To  minimize the contributions of  the unwanted hadronic states and the higher twist dimension operators, we can apply the following power counting scheme,
\begin{align}\label{powercounting}
	\omega_s \sim & \, \omega_M \sim \lambda \, m_{\Lambda_b} \,, & \tilde{\psi}_4 (\omega) \sim & \, \omega .
\end{align}
So the $\Lambda_b \to p $ form factors scales as $1/(n \cdot p')^3 \sim 1/(E_p)^3 $ in the large energy limit of proton, in agreement with the discussion of \cite{mannel_heavy--light_2011,khodjamirian_form_2011}.



\section{Factorization of the correction function at $\mathcal{O}(\alpha_s)$} \label{sect3}
\begin{figure}[htbp]
			\centering
		
			\subfigure[]
			{\label{loopa}
				\begin{minipage}[b]{.3\linewidth}
					\centering
					\includegraphics[scale=0.8]{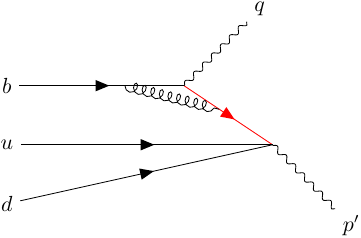}
				\end{minipage}
			}
			\subfigure[]
			{\label{loopb}
				\begin{minipage}[b]{.3\linewidth}
					\centering
					\includegraphics[scale=0.8]{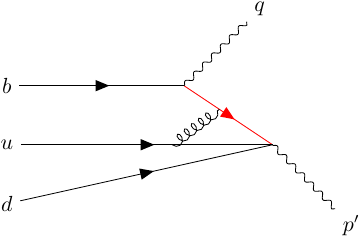}
				\end{minipage}
			}
			\subfigure[]
			{\label{loopc}
				\begin{minipage}[b]{.3\linewidth}
					\centering
					\includegraphics[scale=0.8]{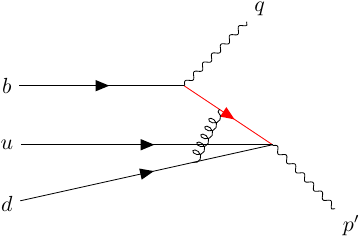}
				\end{minipage}
			} 
				\subfigure[]
			{\label{loopg}
				\begin{minipage}[b]{.3\linewidth}
					\centering
					\includegraphics[scale=0.8]{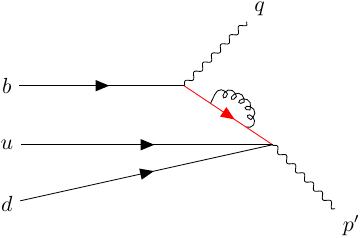}
				\end{minipage}
			}
			\subfigure[]
			{\label{loopd}
				\begin{minipage}[b]{.3\linewidth}
					\centering
					\includegraphics[scale=0.8]{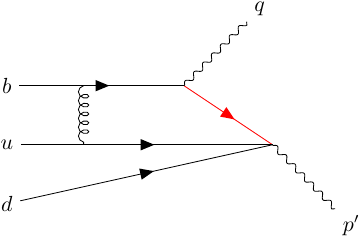}
				\end{minipage}
			}
			\subfigure[]
			{\label{loope}
				\begin{minipage}[b]{.3\linewidth}
					\centering
					\includegraphics[scale=0.8]{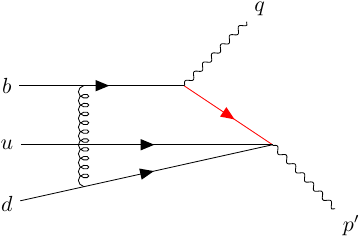}
				\end{minipage}
			}
			\subfigure[]
			{\label{loopf}
				\begin{minipage}[b]{.3\linewidth}
					\centering
					\includegraphics[scale=0.8]{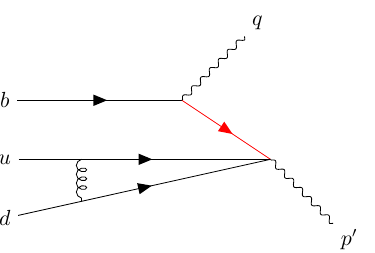}
				\end{minipage}
			}
	
	\caption{Diagrammatical representation of the correlation function $\Pi_{\mu,a}(n\cdot p',\bar{n} \cdot p')$ at one loop. }
	\label{oneloop}	\end{figure}

Because the light-cone operator-product-expansion is valid, when we consider the QCD radiative corrections, the correlation function can be decomposed to a set of independent operators like  \cite{wang_perturbative_2016}
\begin{eqnarray}
  i\int d^4 x e^{ip' \cdot  x } \, \bra{0} T\left\{j_p(x), j_{\mu,a }(0)  \right\} \ket{\Lambda_b(v)} = \sum_{i} \tilde{T}_{i} \otimes\left \langle   \mathcal{O}_{i,\mu} \right \rangle ,
 \end{eqnarray}
where $\otimes$ denotes a convolution in the light-cone variables $\omega_1$ and $\omega_2$.
 Whenever necessary, the independent operators $\mathcal{O}_{i}$ should include the physical operators and the "evanescent operators". Fortunately, our calculations demonstrate that physical operators do not mix with evanescent operators at NLO accuracy. Consequently, the contributions from evanescent operators can be safely neglected.

The hard kernel $\tilde{T}_{i}$ consists of the hard function $C_{i}$ and the jet function $J_{i}$,
\begin{eqnarray}
     \tilde{T}_{i}  =  C_{i} \cdot J_{i} .
\end{eqnarray}
We will employ the method of regions to evaluate the hard coefficients and the jet
 functions simultaneously. 
 The hard function is from the sum of all diagrams in hard region and the hard-collinear contributions of the diagrams lead to the jet function. The sum of the soft contributions is identical to the NLO partonic distribution amplitude with the tree-level hard kernel, which is absorbed by the matrix element of the operators $\left \langle   \mathcal{O}_{i,\mu} \right \rangle$ . 
 So we just consider the hard and hard-collinear contributions when calculating the NLO hard kernel.

Though we seem to face operator mixing after computing the loop integrals at NLO accuracy, these mixed operators can be reduced to a combination involving a twist-4 LCDA and a set of twist-3 LCDAs for the $\Lambda_b$ baryon, through applying the momentum-space light-cone projector of the $\Lambda_b $-baryon.
 The isospin invariance for the light-quark fields in $\Lambda_b$-baryon can lead to the cancellation of twist-3 LCDAs. So like tree-level factorization formulae, only one twist-4 LCDA of the $\Lambda_b$ baryon entering NLO factorization formulae. Then the ultraviolet(UV) and infrared(IR) subtractions can be conducted  without the operator mixing.
Finally we can have the factorized expressions of the correlation functions,
\begin{eqnarray}
	 \Pi_{\mu,a }(P,q) =  \tilde{T}_{b}  \otimes \Phi_{\Lambda_b, b} =  C_{b} \cdot J_{b} \otimes \Phi_{\Lambda_b ,b } = Z_C C_{R}  \cdot  Z_J J_{R} \otimes Z_{\Phi}\Phi_{\Lambda_b ,R },
\end{eqnarray}
where the subscripts $b$ and $R$ denote the bare and renormalized operators, respectively, and $ \Phi_{\Lambda_b}$ is the distribution amplitude of $\Lambda_b-$baryon. $Z_C\,,\,Z_J$ and $Z_{\Phi} $ are the renormalization constant for the  hard function , jet function and distribution amplitude.
By computing convolution integrals of the NLO partonic distribution amplitude with
 the tree-level hard kernel,  we will verify that $Z_C Z_J Z_{\Phi}$ can cancel the the factorization-scale dependence completely in the factorized expressions of the  correlation functions $\Pi_{\mu,a }(P,q)$ and the residual scale dependence  stems from the UV renormalization of
the baryonic current as displayed in \Eq{RGfn}.Resummation of large logarithms involved in the
 perturbative functions is carried out at NLL using the momentum-space RG approach.

 Because Fig.\ref{loopg} only consists of the one-gluon exchange between the two soft quarks, it only provides the soft contributions and no contribution to the perturbative
functions can arise from this diagram. So we just compute  other one-loop QCD diagrams displayed in Fig.\ref{oneloop} for determinations of perturbative
functions.

So the contributions of all the one-loop QCD diagrams on the correlation function can be written as 
\begin{eqnarray}\label{phibud}
	 \Phi^{(0)}_{\Lambda_b} \otimes T^{(1)}_{V(A)} =  \Sigma_{i \in \left\{a,b,c,d,e,f\right\}}\Pi^{i}_{\mu, V(A) }(P,q) \, ,
\end{eqnarray}
where $ \Pi^{i }_{\mu, V(A) }(P,q) $ is the one-loop QCD correction from the corresponding diagram in Fig.\ref{oneloop}.

\subsection{Weak vertex diagram}

 Now we turn to compute the one-loop QCD correction to  the partonic correlation function $\Pi_{\mu,a}^{{\rm par}}$ from the weak vertex diagram displayed in Fig. \ref{loopa}

 \begin{eqnarray}\label{eqfiga}
    \Pi^{{\rm par} ,a }_{\mu, V(A) }(P,q) &=& \frac{ 2 i g^2_s C_F }{ \bar{n} \cdot p' - \omega_1 - \omega_2}  \int  \frac{{\d}^D \ell }{ (2 \pi)^D} \frac{1 }{[(p'-k+ \ell)^2 +i0 ] [(m_b v + \ell)^2 - m_b^2 + i0 ] [\ell^2 + i0 ]  }  \notag\\
	&&  [u^T(k_1) C  \not{n}  \frac{\not{\bar{n}}}{2} \gamma_{\rho} (\not{p}' - \not{k} + \not{\ell }) \gamma_{\mu} (1, \gamma_5)  (m_b \not{v} +\not{\ell }  +m_b  )  \gamma^{\rho} b(v)  ]  \gamma_5 \not{n} {d}(k_2) \, ,    
 \end{eqnarray}
where $k= k_1 + k_2 \,,\, \omega_i = \bar{n} \cdot k_i \, (i=1,2)  $ and $D= 4-2 \varepsilon$. 
The factor two originates from the two $u$-quark possess exchange symmetry. 
We have approximated the $b$-quark momentum as $p_b=m_b v$ by dropping out the residual momentum, since we are only interested in extracting the leading power contributions to the correlation functions.


By expanding \Eq{eqfiga} in the hard region and applying the light-cone projector subsequently, the hard function contributed from Fig.\ref{loopa} can be determined,

\begin{eqnarray}
    \Pi^{a,h}_{\mu, V(A) }(P,q) &=&\frac{ig^2_s C_F f^{(2)}_{\Lambda_b}(\mu) \phi_4(\omega_1,\omega_2)  }{ \bar{n} \cdot p' - \omega_1 - \omega_2)}  \int  \frac{{d}^D {\ell} }{ (2 \pi)^D}  \frac{(1,-\gamma_5) }{[(p' + \ell )^2 +i0 ] [(m_b v +{\ell})^2 - m_b^2 + i0 ] [{\ell}^2 + i0 ]  }  \notag\\
	&& \times \left\{ \gamma_{\perp \mu} \big[ 2 n \cdot {\ell} (\bar{n} \cdot {\ell} + m_b  ) + 2 n \cdot p' (\bar{n} \cdot {\ell} +m_b) + (D-4){\ell}^2 \big] \right .  \notag\\
	&& \left . + n_{\mu} (2-D)(\bar{n} \cdot {\ell})^2  +{\bar{n}}_{\mu} \big[2m_b({n} \cdot p' + n \cdot {\ell} ) + (D-2) {\ell}^2_{\perp} \big] \right \}  \, \slashed n u_{\Lambda_b}(P) \, ,
\end{eqnarray}
 where the superscript "par" of the partonic correlation functions $ \Pi^{a,h}_{\mu, V(A)}$ is suppressed from
 now on and we have introduced
\begin{align}
    \ell^2_{\perp} \equiv & \, g^{\mu \nu}_{\perp} \ell_{\mu} \ell_{\nu}  \,, &  g^{\mu \nu}_{\perp} = & g^{\mu \nu} - \frac{ n^{\mu} \bar{n}^{\nu}}{2 } -\frac{ n^{\nu} \bar{n}^{\mu}}{2 } \, .  
\end{align}

Evaluating the loop integrals, one attain 
\begin{align}
     \Pi^{a,h}_{\mu, V(A) }(P,q) = & \frac{\alpha_s C_F}{4 \pi}  \frac{ f^{(2)}_{\Lambda_b}(\mu) \, \phi_4(\omega_1,\omega_2) }{\bar{n} \cdot p' - \omega_1 - \omega_2 } (1,-\gamma_5) \left[\gamma_{\perp\mu} C^{(a)}_{h,\perp}(n\cdot p')  +n_{\mu} C^{(a)}_{h,n}(n\cdot p') \right .
    \nonumber \\
    & \left . + {\bar{n}}_{\mu} C^{(a)}_{h,\bar{n}}(n\cdot p')    \right] \, \slashed n u_{\Lambda_b}(P) \, ,
\end{align}
where the coefficients functions read 
\begin{eqnarray}
    C^{(a)}_{h,\perp}(n\cdot p') &=& \left ( \frac{1}{\varepsilon} + 1 \right ) \left ( {1 \over \epsilon} + 2  \ln { \mu \over n \cdot p'} \right )  + 2  \ln^2 \frac{ \mu}{n \cdot p'} + 2{\rm Li_2} \left( 1 - r \right) + \frac{r}{1-r} \ln r + \frac{\pi^2}{12} + 4 \, ,
\nonumber \\
    C^{(a)}_{h,\bar{n}}(n\cdot p') &=& \left ( \frac{1}{\varepsilon} + 1 \right ) \left ( {1 \over \epsilon} + 2  \ln { \mu \over n \cdot p'} \right ) + 2\ln^2 \frac{ \mu}{n \cdot p'} + 2{\rm Li_2} \left( 1 - r \right) + \frac{r}{r-1} \ln r + \frac{\pi^2}{12} +3 \, ,
\nonumber \\
    C^{(a)}_{h,n} (n\cdot p') &=& \frac{1}{r-1} \left( 1+ \frac{r}{1-r} \ln r \right) \, ,
\end{eqnarray}
and $r= n \cdot p' / m_b$.

 By proceeding in a similar way, we can extract the hard-collinear contribution from Fig\ref{loopa} as follows
 \begin{align}
    \Pi^{a,hc}_{\mu, V(A) }(P,q)  = & \frac{ig^2_s C_F f^{(2)}_{\Lambda_b}(\mu) \phi_4(\omega_1,\omega_2)  }{ \bar{n} \cdot p' - \omega_1 - \omega_2} \,\, \left [ (1,-\gamma_5)  \left ( \gamma_{\perp \mu}   +{\bar{n}}_{\mu} \right ) \, \slashed n u_{\Lambda_b}(P) \right ]  \notag\\
	& \times \int  \frac{{\d}^D {\ell} }{ (2 \pi)^D} \frac{2 m_b n \cdot (p'+{\ell}) }{[\bar{n } \cdot (p'-k+{\ell}) n \cdot (p'+{\ell}) +{\ell}^2_{\perp} + i0 ] [m_b  n\cdot {\ell}  + i0 ] [{\ell}^2 + i0 ]  } \, .
 \end{align}
Then we write 
 \begin{align}
    \Pi^{a,hc}_{\mu, V(A) }(P,q) = & - \frac{\alpha_s C_F }{4\pi}\frac{f^{(2)}_{\Lambda_b}(\mu) \phi_4(\omega_1,\omega_2)  }{ \bar{n} \cdot p' - \omega_1 - \omega_2} \, \left [ (1,-\gamma_5)  \left ( \gamma_{\perp \mu}   + {\bar{n}}_\mu \right )  \, \slashed n u_{\Lambda_b}(P)\right ] 
\nonumber \\
	& \left [ \left ( \frac{2}{\varepsilon} + 2 \right ) \left( {1 \over \epsilon} + \ln \frac{\mu^2}{n \cdot p' (\omega - \bar{n} \cdot p' )} \right)  +\ln^2 \frac{\mu^2}{n \cdot p' (\omega - \bar{n} \cdot p' )} - \frac{\pi^2}{6} + 4 \right ] ,
\end{align}
where $\omega =\omega_1 + \omega_2$.

\subsection{Proton vertex diagrams}
The one-loop contributions to $\Pi_{\mu ,V(A)}(P,q) $ from the proton vertex diagrams shown in Fig.\ref{loopb}  and Fig.\ref{loopc} are given by 
\begin{eqnarray}\label{eqfigb}
    \Pi^b_{\mu, V(A)} (P,q ) &=& { - 2 i g_s^2 C_F \over N_c - 1} \frac{1}{ (p' - k)^2  }  \int \frac{d^D {\ell} }{(2\pi)^D}  \frac{1}{[(p'-k_2 -{\ell} )^2 + i0] [( \ell - k_1)^2 + i0] [\ell^2 + i0 ]} \notag \\
	&& [u^T(k_1) C \gamma_{\rho} \not{{\ell}} \not{n} (\not{p}' - \not{k_2} -\not{{\ell}}  )  \gamma^{\rho} (\not{p}' - \not{k_1} -\not{k_2}  )   \gamma_{\mu} (1, \gamma_5) b(v)   ] \gamma_5 \not{n} d (k_2) ,  
\end{eqnarray}
\begin{eqnarray}\label{eqfigc}
    \Pi^c_{\mu, V(A)} ( P,q ) &=& { - 2 i g_s^2 C_F \over N_c - 1} \frac{1}{ (p' - k)^2  }  \int \frac{d^D {\ell} }{(2\pi)^D}  \frac{1}{[(p'-k_1 -{\ell} )^2 + i0] [( \ell - k_2)^2 + i0] [\ell^2 + i0 ]} \notag \\
	&& [u^T(k_1) C   \not{n} (\not{p}' - \not{k_1} -\not{{\ell}}  )  \gamma^{\rho} (\not{p}' - \not{k_1} -\not{k_2}  )   \gamma_{\mu} (1, \gamma_5) b(v)    ]  \gamma_5 \not{n}   \not{l}  \gamma_{\rho}  d(k_2)  \, . 
\end{eqnarray}
It is more convenient to calculate the loop integral exactly. Then we  keep only the leading power terms in the
 resulting partonic amplitude and insert the light-cone projector of the $\Lambda_b$-baryon.   Based on the argument from the power counting analysis about the loop integral, the leading power contribution to $\Pi^b_{\mu, V(A)}$, $\Pi^c_{\mu, V(A)}$  only arise from the hard-collinear region and the resulting contribution to the jet function is found to be 
 \begin{eqnarray}
    \Pi^{b,hc}_{\mu, V(A)} &=& {- \alpha_s C_F \over 2 \pi (N_c - 1)} \frac{ f^{(2)}_{\Lambda_b}(\mu) \phi_4(\omega_1,\omega_2) }{\bar{n} \cdot p' - \omega_1 - \omega_2 }   \, \left [ (1,- \gamma_5) \left ( \gamma_{\mu\perp} + \bar n_\mu \right ) \not{n} u_{\Lambda_b}(P) \right ]  \notag\\
	&& \left [ \left ( \frac{ 1+\eta_2}{\eta_1} \ln \frac{1+ \eta_{12} }{1+ \eta_2} - {1 \over 2} \right )    \left ( \frac{1}{\varepsilon} + \ln \frac{\mu^2}{n \cdot p' (\omega_2 - \bar{n} \cdot p') } \right ) -\frac{ 1+\eta_2}{2\eta_1} \ln^2 \frac{1+\eta_{12}}{1+\eta_2}  \right.\notag\\
    &&\left.+ \frac{ 2(1+\eta_2) + \eta_1}{\eta_1} \ln \frac{1+\eta_{12}}{1+\eta_2}  -2  \right ] \notag\\
    &&+{ \alpha_s C_F \over 2 \pi (N_c - 1)} \frac{ f^{(2)}_{\Lambda_b}(\mu) \phi_3^{+-}(\omega_1,\omega_2)}{\bar{n} \cdot p' - \omega_1 - \omega_2 }    \, [(1,\gamma_5)  \frac{\not{n} }{4}  \gamma_{\mu}u_{\Lambda_b}(P) ]   \frac{ 1+\eta_{12}}{\eta_1}\ln \frac{1+ \eta_{12} }{1+ \eta_2},
 \end{eqnarray} 
 \begin{eqnarray}
    \Pi^{c,hc}_{\mu, V(A)} &=&{- \alpha_s C_F \over 2 \pi (N_c - 1)} \frac{ f^{(2)}_{\Lambda_b}(\mu) \phi_4(\omega_1,\omega_2) }{\bar{n} \cdot p' - \omega_1 - \omega_2 }   \, \left [ (1,- \gamma_5) \left ( \gamma_{\mu\perp} + \bar n_\mu \right ) \not{n} u_{\Lambda_b}(P) \right ]   \notag\\
   &&  \left [ \left( \frac{ 1+\eta_1}{\eta_2}\ln \frac{1+ \eta_{12} }{1+ \eta_1} - 1\right)  \left(\frac{1}{\varepsilon} + \ln \frac{\mu^2}{n \cdot p' (\omega_1 - \bar{n} \cdot p') } \right) -\frac{ 1+\eta_1}{2\eta_2} \ln^2 \frac{1+\eta_{12}}{1+\eta_1}  \right.\notag\\
    &&\left. + \frac{ 2(1+\eta_1) + \eta_2}{\eta_2} \ln \frac{1+\eta_{12}}{1+\eta_1}  -\frac{3}{2}   \right] \notag\\
    &&-{ \alpha_s C_F \over 2 \pi (N_c - 1)} \frac{ f^{(2)}_{\Lambda_b}(\mu) \phi_3^{-+}(\omega_1,\omega_2)}{\bar{n} \cdot p' - \omega_1 - \omega_2  }    \, [(1,\gamma_5)  \frac{\not{n} }{4}  \gamma_{\mu}u_{\Lambda_b}(P) ]    \frac{ 1+\eta_{12}}{\eta_2}\ln \frac{1+ \eta_{12} }{1+ \eta_1},
 \end{eqnarray} 
where    we define
\begin{align}
    \eta_i = & - {\omega_i \over \bar{n} \cdot p'} \,, & \eta_{12} = & \eta_1 +\eta_2.
\end{align}

$ \phi_3^{+-}(\omega_1,\omega_2)  ,  \phi_3^{-+}(\omega_2,\omega_1)$ are the twist-3 LCDAs of $\Lambda_b-$baryon and they satisfy the relation that $ \phi_3^{+-}(\omega_1,\omega_2)  =  \phi_3^{-+}(\omega_2,\omega_1)$  \cite{bell_light-cone_2013} . 
The contribution to the jet function of $\phi_3^{+-(-+)}$ from diagram $b$ and $c$ are symmetric under the exchange $\omega_1 \leftrightarrow \omega_2$ ,
the terms with $ \phi_3^{+-(-+)}(\omega_1,\omega_2)$ can be eliminated after summing the corrections from the two proton vertex diagrams.

\subsection{Wave function renormalization}
For the wave fucntion renormalization, the self-energy correction to the intermediate quark propagator in Fig\ref{loopd} has  the hard-collinear contribution. It is independent of the Dirac structures of the weak transition current operator and the baryonic interpolating current operator. It is straightforward to write 
\begin{eqnarray}
    \Pi^{d,hc}_{\mu,V(A)} (P,q) &=& \frac{\alpha_s C_F}{4 \pi} \, \frac{f^{(2)}_{\Lambda_b}(\mu) \phi_4(\omega_1,\omega_2) }{ \bar{n} \cdot p' - \omega_1 - \omega_2   } \, \left [ (1,- \gamma_5) \left ( \gamma_{\perp\mu}+ \bar{n}_{\mu} \right ) \, \slashed n u_{\Lambda_b}(P) \right ] \notag\\
	&& \times \left ( \frac{1}{\varepsilon} + \ln \frac{\mu^2}{n \cdot p' (\omega - \bar{n} \cdot p') }   + 1 \right ) \, .
\end{eqnarray}

The contributions of the wave function renormalization to the external quark fields can be taken from 
\begin{eqnarray}
     \Pi^{bwf,(1)}_{\mu,V(A)} - \Phi^{(1)}_{\Lambda_b,bwf} \otimes T^{(0)} & =& \frac{\alpha_s C_F}{8 \pi} \frac{f^{(2)}_{\Lambda_b}(\mu) \phi_4(\omega_1,\omega_2) }{  \bar{n} \cdot p' - \omega_1 - \omega_2  }   \left [ (1,- \gamma_5) \left ( \gamma_{\perp\mu}+ \bar{n}_{\mu} \right ) \, \slashed n u_{\Lambda_b}(P) \right ]  \notag\\
	 && \times \left ( \frac{3}{\varepsilon} + 3\ln \frac{\mu^2}{m_b^2 }   +4 \right ) \, ,
\nonumber \\
    \Pi^{uwf,(1)}_{\mu,V(A)} - \Phi^{(1)}_{\Lambda_b,uwf} \otimes T^{(0)} & = &    \Pi^{dwf,(1)}_{\mu,V(A)} - \Phi^{(1)}_{\Lambda_b,dwf} \otimes T^{(0)} =0 .
\end{eqnarray}
where $\Pi^{qwf,(1)}_{\mu,V(A)} (q=b,u,d) $ stands for the contribution to $\Pi_{\mu,V(A)}$ from the wave function renormalization of the $q$-quark field at one loop, and $\Phi^{(1)}_{\Lambda_b,qwf} $ denotes the one-loop contribution to the distribution amplitude of $\Lambda_b-$baryon from  field renormalization of the $q$-quark.

\subsection{Box diagrams}

We proceed to compute the one-loop contribution from the two box diagrams displayed in Fig.\ref{loope} and Fig.\ref{loopf} . We can readily write 
\begin{eqnarray}
    \Pi^e_{\mu,V(A)} (P,q) &=& -ig_s^2 {2 C_F \over N_c - 1} \int \frac{d^D {\ell} }{(2\pi)^D}  \frac{1}{[(p'-k+{\ell} )^2+ i0] [({\ell}-k_1)^2 + i0]  [(m_b v +{\ell})^2-m_b^2 + i0 ]}  
\notag \\
	&& \times {1 \over [\ell^2 + i0]} [u^T(k_1) C \gamma_{\rho} (\not{k}_1- \not{{\ell}}) \not{n} (\not{p}' - \not{k} +\not{{\ell}}  )  \gamma_{\mu} (1, \gamma_5)   (m_b \not{v} + \not{{\ell}} +m_b) \gamma^{\rho}    b(v)    ]
\nonumber \\
    &&  \times \gamma_5 \slashed n \, d(k_2),
\label{eqfige} \\
    \Pi^f_{\mu, V(A)} ( P,q ) &=& -ig_s^2 {2 C_F \over N_c - 1} \int \frac{d^D {\ell} }{(2\pi)^D}  \frac{1}{[(p'-k+{\ell} )^2+ i0] [({\ell}-k_2)^2 + i0]  [(m_b v +{\ell})^2-m_b^2 + i0 ] }
\notag\\
	&& \times {1 \over [\ell^2 + i0]} [u^T(k_1) C   \not{n} (\not{p}' - \not{k} +\not{{\ell}}  )  \gamma_{\mu} (1, \gamma_5)   (m_b \not{v} + \not{{\ell}} +m_b) \gamma^{\rho}    b(v)    ] \notag\\
	&&  \times\gamma_5 \not{n}   (\not{k}_2-\not{l})  \gamma_{\rho}  \, d(k_2) \, . \label{eqfigf}
\end{eqnarray}
 It is evident that the  hard contribution is power suppressed and the contribution 
 to the jet function  can arise from the box diagrams, Fig. \ref{loope} and Fig. \ref{loopf}. They can be determined by expanding \Eq{eqfige} and \Eq{eqfigf} in the hard-collinear region systematically,

 \begin{eqnarray}
    \Pi^{e,hc}_{\mu, V(A)} &=& - {\alpha_s \over 4\pi}  \, {C_F \over N_c - 1} \, \frac{ 1 }{\omega_1 } \left [ (1,- \gamma_5) \left ( \gamma_{\perp\mu}+ \bar{n}_{\mu} \right ) \, \slashed n u_{\Lambda_b}(P) \right ]  \, \ln \frac{1+ \eta_{12} }{1+ \eta_2} \, \Big [ f^{(2)}_{\Lambda_b}(\mu) \phi_4(\omega_1,\omega_2) \notag\\
	&& \times \left ( \frac{2}{\varepsilon} + \ln \frac{\mu^2}{n \cdot p' (\omega - \bar{n} \cdot p') } + \ln \frac{1+ \eta_{12} }{1+ \eta_2} + 2  \right ) +  {n \cdot p' \over 2 m_b} \,  f^{(1)}_{\Lambda_b}(\mu) \phi_3^{+-}(\omega_1,\omega_2) \, \Big ] \, , \notag\\
\nonumber \\
    \Pi^{f,hc}_{\mu, V(A)} &=&  - {\alpha_s \over 4\pi}  \, {C_F \over N_c - 1} \, \frac{1}{\omega_2 } \left [ (1,- \gamma_5) \left ( \gamma_{\perp\mu}+ \bar{n}_{\mu} \right ) \, \slashed n u_{\Lambda_b}(P) \right ]   \, \ln \frac{1+ \eta_{12} }{1+ \eta_1} \Big [  f^{(2)}_{\Lambda_b}(\mu) \phi_4(\omega_1,\omega_2) \notag\\
	&& \times \left ( \frac{2}{\varepsilon} + \ln \frac{\mu^2}{n \cdot p' (\omega - \bar{n} \cdot p') } + \ln \frac{1+ \eta_{12} }{1+ \eta_1}+ 2 \right ) - \, {n \cdot p' \over 2 m_b}  f^{(1)}_{\Lambda_b}(\mu) \phi_3^{-+}(\omega_1,\omega_2)  \, \Big ] \, .
 \end{eqnarray} 

Although terms with twist-3 LCDAs appear, they still end up being canceled by a similar analysis of the proton vertex diagrams.

\subsection{ The NLO hard-scattering kernels}
We are ready to determine the one-loop hard kernels entering QCD factorization formulae
 of the correlation functions $\Pi^{\rm par}_{\mu ,V(A)} (P,q)$ by collecting different pieces together 
\begin{eqnarray}
    \Phi^{(0)}_{\Lambda_b} \otimes T^{(1)}_{V(A)} &=& \int d \omega_1 \, d \omega_2 \biggl\{\left[\Pi^{a,h}_{\mu, V(A)}  + (\Pi^{bwf,(1)}_{\mu,V(A)}  - \Phi^{(1)}_{\Lambda_b,bwf} \otimes T^{(0)}_{V(A)}  ) \right] \notag\\
&&+\left[ \Pi^{a,hc}_{\mu, V(A)} +\Pi^{b,hc}_{\mu, V(A)} +\Pi^{c,hc}_{\mu, V(A)} +\Pi^{d,hc}_{\mu, V(A)} +\Pi^{e,hc}_{\mu, V(A)} +\Pi^{f,hc}_{\mu, V(A)}\right]\biggr\} \, ,
\end{eqnarray}
where the terms in the first and second square brackets correspond to the hard and jet functions at $\mathcal{O}(\alpha_s)$, respectively. Introducing the definition, 
\begin{eqnarray}
    \Pi_{\mu,V(A)} = (1,-\gamma_5) \left[  \Pi_{\perp,V(A)}  \gamma_{\perp \mu} + \Pi_{\bar{n} ,V(A)} \bar{n}_{\mu}  +\Pi_{{n} ,V(A)} {n}_{\mu}  \right] \slashed n \, u_{\Lambda_b}(P) \, ,
\end{eqnarray}
 we can readily obtain the following factorization formulae for the vacuum-to-$\Lambda_b$-baryon correlation functions at NLO
\begin{eqnarray}
    \Pi_{m,V(A)} &=& f^{(2)}_{\Lambda_b}(\mu) \, C_{m,V(A)}(n \cdot p' ,\mu) \, \int_{0}^{\infty} d\omega_1 \, d\omega_2 \, \frac{\phi_4(\omega_1,\omega_2, \mu)}{\omega_1 + \omega_2 - \bar{n} \cdot p' - i0 } \, J_{m} \left ( \bar n \cdot p' , \omega_i, \mu \right ) \, ,
\end{eqnarray}
where $m \in \{ \perp \, , n \, , \bar n \}$ and the renormalized hard coefficients are given by 
\begin{eqnarray}
    C_{\bar{n},V(A)}(n \cdot p' ,\mu) &=& 1- \frac{\alpha_s(\mu) C_F}{4 \pi } \left [ 2 \ln^2 \frac{\mu}{n \cdot p'}  +5\ln \frac{\mu}{m_b} + 2{\rm Li_2} \left( 1 - r \right)  + \frac{r-2}{1-r} \ln r + \frac{\pi^2 }{12} +5 \right ] \, ,
\nonumber \\
    C_{\perp,V(A)}(n \cdot p' ,\mu) &=& C_{\bar{n},V(A)}(n \cdot p' ,\mu) - \frac{\alpha_s(\mu) C_F}{4 \pi } \left ( 1 + {2 r \over 1 - r} \ln r \right ) \, ,
\nonumber \\
    C_{n,V(A)}(n \cdot p' ,\mu) &=& - \frac{\alpha_s(\mu) C_F}{4 \pi } \left [\frac{1}{r-1} \left(1+\frac{r}{1-r} \ln r \right) \right ] \, ,
 \end{eqnarray}
and the renormalized jet functions read
\begin{eqnarray}
    J_{\perp(\bar n)} \left ( \bar n \cdot p' , \omega_i, \mu \right ) &=& 1 - \frac{ \alpha_s (\mu) }{3 \pi} \left [  \ln^2  \frac{\omega- \bar{n} \cdot p'}{ \omega_2- \bar{n} \cdot p'} - 2 \ln  \frac{\omega- \bar{n} \cdot p'}{ \omega_2- \bar{n} \cdot p'}  \left ( \frac{\omega_2- \bar{n} \cdot p'}{ \omega_1} -\frac{3}{4} \right ) + \frac{\pi^2}{6} + \frac{1}{2}  \right .
\notag\\
	&& \left . - \ln \frac{\mu^2}{n \cdot p' (\omega- \bar{n} \cdot p')} \left ( \ln \frac{\mu^2}{n \cdot p' (\omega- \bar{n} \cdot p')} - 2 \ln  \frac{\omega- \bar{n} \cdot p'}{ \omega_2- \bar{n} \cdot p'} - {1 \over 2}\right )\right ] \, ,
\nonumber \\
    J_{n} \left ( \bar n \cdot p' , \omega_i, \mu \right ) &= & 1 + \mathcal O(\alpha_s) \, .
\end{eqnarray}
Surprisingly, the jet and hard functions are identical to their counterparts in the $\Lambda_b \to \Lambda$ process \cite{wang_perturbative_2016}. 
This similarity can be traced back to the interpolating currents.  In \cite{wang_perturbative_2016} and our work, the leading power current operators of  the $\Lambda$-baryon and  the proton are employed, respectively, which are not suppressed at leading power. 
Given that the proton and the $\Lambda$ baryon belong to the same SU(3) flavor octet and quark mass effects are neglected, flavor symmetry dictates that these currents are analogous. This underlying analogy likely gives rise to the identical jet and hard functions.

The hard coefficients we calculate are in consistent to the relations $C_{\perp,V }=C_{\perp,A } \,,\, C_{\bar{n},V }=C_{\bar{n},A }\,,\, C_{n,V }=C_{n,A }$ to all orders in perturbative theory due to the $\rm U(1)$ helicity symmetry for both massless QCD and SCET Lagrangian functions \cite{PhysRevD.63.114020}.
Therefore, any difference between the axial-vector and vector 
$\Lambda_b \to p $ form factors must arise from the jet function. 
Obviously, at large hadronic recoil, the  axial-vector and vector form factors are identical at leading power in $\lambda$.

For the RG evolution equations of the correlation fucntion, it is
\begin{eqnarray}
    \frac{d \, \Pi_{\perp(\bar n),V(A)}  }{d \,\ln \mu }	&=& \frac{\alpha_s(\mu)}{3\pi} \int_{0}^{\infty} d \omega_1 d \omega_2 \frac{f^{(2)}_{\Lambda_b}(\mu) \phi_4( \omega_1,\omega_2 ,\mu) }{\omega_1 + \omega_2 - \bar{n} \cdot p' - i0} \left[4 \ln \frac{\mu}{\omega-\bar{n} \cdot p'}  -4 \ln \frac{\omega-\bar{n}\cdot p' }{\omega_2-\bar{n} \cdot p'} -6 \right]  \notag\\
    && + \int_{0}^{\infty} d \omega_1 d \omega_2 \frac{1 }{\omega_1 + \omega_2 - \bar{n} \cdot p' - i0} \frac{d}{d \ln \, \mu} \left[f^{(2)}_{\Lambda_b}(\mu) \phi_4( \omega_1,\omega_2 ,\mu)\right] + \mathcal{O}(\alpha_s^2) \, ,
\nonumber \\
	\frac{d \, \Pi_{n,V(A)}  }{d \, \ln \mu } &=&  \mathcal{O}(\alpha_s^2).
\end{eqnarray}

The unknown one-loop evolution equation for $f^{(2)}_{\Lambda_b}(\mu) \phi_4(\omega_1,\omega_2,\mu)$ can be determined from the NLO partonic distribution amplitudes with the tree-level hard kernel.
 The diagrams for these NLO partonic distribution amplitudes are identical between the $\Lambda_b \to p$ and $\Lambda_b \to \Lambda$  processes \cite{wang_perturbative_2016} . After calculation, we conclude that the evolution equation is as follows,
\begin{eqnarray}
&&\int_{0}^{\infty} d\omega_1 d\omega_2 \frac{1 }{\omega_1 + \omega_2 - \bar{n} \cdot p' - i0} \frac{d}{d \, \ln \mu} \left[f^{(2)}_{\Lambda_b}(\mu) \phi_4( \omega_1,\omega_2 ,\mu)\right] \notag\\
&&=-\frac{\alpha_s(\mu)}{3\pi} \int_{0}^{\infty} d\omega_1 d\omega_2 \frac{f^{(2)}_{\Lambda_b}(\mu) \phi_4( \omega_1,\omega_2 ,\mu) }{\omega_1 + \omega_2 - \bar{n} \cdot p' - i0} \left[4 \ln \frac{\mu}{\omega-\bar{n} \cdot p'}  -4 \ln \frac{\omega-\bar{n}\cdot p' }{\omega_2-\bar{n} \cdot p'} -5 \right]  \, .
\end{eqnarray}
So the correlation function $\Pi_{\mu,a} (P,q)$ satisfies 
\begin{eqnarray}
	\frac{d \,\Pi_{\perp(\bar{n}),V(A)}  }{d \, \ln \mu }	= -\frac{\alpha_s(\mu)}{3\pi} \int_{0}^{\infty} d \omega_1 d \omega_2 \frac{ f^{(2)}_{\Lambda_b}(\mu) \phi_4( \omega_1,\omega_2 ,\mu)  }{\omega_1 + \omega_2 - \bar{n} \cdot p' - i0} =-\frac{\alpha_s (\mu)}{4 \pi} \frac{4}{3} \Pi_{\perp(\bar{n}),V(A)}  + \mathcal{O}(\alpha_s^2).
\end{eqnarray}
This residual $\mu$-dependence of $\Pi_{\perp(\bar{n}),V(A)} $ can cancel against that from the interpolating current operator, $f_N(\mu)$ in \Eq{Interpolating}. So all 6 form factors  has the factorization-scale independence at one loop.

\subsection{Resummation of large logarithms}
We designate the factorization scale $\mu$  as a hard-collinear scale $\mu_{\rm hc}$ of order $\lambda^{1/2} m_{\Lambda_b}$ , and it is comparable to the hadronic  scale $\mu_0 \simeq 1 {\rm GeV}$  entering the initial condition of the 
$\Lambda_b-$baryon LCDAs. So we will not resum logarithms of $\mu / \mu_0$. However, the large logarithms of $\mu / m_b$ in the hard functions can not be avoided. We have to solve RG equations in momentum space to resum them at NLL.

The  RG evolution equations of the hard functions can been written as 
 \begin{eqnarray}
    \frac{d}{d\, \ln \mu} C_i (n \cdot p' ,\mu ,\nu') = \left[ - \Gamma_{\rm cusp} \ln \frac{\mu}{n \cdot p'}  + \gamma(\alpha_s)\right]  C_i (n \cdot p' ,\mu ,\nu'),
 \end{eqnarray}
where $C_i$ stands for $ C_{\perp,V(A)} , C_{\bar{n},V(A)}, C_{n,V(A)}  $ and the cusp anomalous dimension $\Gamma_{\rm cusp}(\alpha_s) $ at three-loop order and the remaining anomalous dimensions $\gamma(\alpha_s)$ at two-loop order can be found in \cite{beneke_b_2011} . Then we have 
\begin{eqnarray}
     C_{\perp(\bar{n}) ,V(A)} (n \cdot p' ,\mu )  = U_1 (\bar{n} \cdot p' ,\mu_h , \mu )  C_{\perp(\bar{n}) ,V(A)} (n \cdot p' ,\mu_h ) \, .
\end{eqnarray}
Here $U_1 (\bar{n} \cdot p',\mu_h , \mu ) $ can be deduced from $U_1 (E_{\gamma},\mu_h , \mu )$ in \cite{beneke_b_2011} with $E_{\gamma} \to \frac{\bar{n} \cdot p'}{2}$. 
 $\mu_h$ should be set to a hard scale of order $ n \cdot p' \sim m_b $. 

In addition, as mentioned in \cite{wang_perturbative_2016}, it is necessary to distinguish the renormalization and the factorization scales, which are set to be equal in dimensional regularization.
So we introduce the renormalization scale $\nu$.  Following \cite{bell_heavy--light_2011,wang_perturbative_2016}, 
the distinction between the
 renormalization and the factorization scales can be accounted by writing 
\begin{eqnarray}
      J_{\perp(\bar n)} \left ( \bar n \cdot p', \omega_i, \mu, \nu \right ) =   J_{\perp(\bar n)} \left ( \bar n \cdot p', \omega_i, \mu \right ) + \Delta J_{\perp(\bar n)} \left ( \bar n \cdot p', \omega_i, \nu \right ) .
\end{eqnarray}
Exploiting the RG evolution equations,
\begin{eqnarray}
    \frac{d}{d \, \ln \nu } \left[ \ln  \Delta   J_{\perp(\bar n)} \left ( \bar n \cdot p', \omega_i, \nu \right )  \right]  = - \sum_k  \left ( \frac{\alpha_s}{4\pi} \right )^k \gamma_{N}^{(k)},
\end{eqnarray}
and implementing the renormalization conditions 
\begin{eqnarray}
     \Delta   J_{\perp(\bar n)} \left ( \bar n \cdot p', \omega_i, \mu \right ) = 0  ,
\end{eqnarray}
we have
\begin{eqnarray}
     \Delta   J_{\perp(\bar n)} \left ( \bar n \cdot p', \omega_i, \nu \right ) = - \frac{\alpha_s }{4\pi}  \gamma_{p}^{(1)} \ln \frac{\nu}{\mu}  +\mathcal{O}(\alpha_s^2) \, ,
\end{eqnarray}
 where $\gamma_{p}^{(1)} = 4/3$ has been defined in \Eq{RGfn}.

Now the jet function becomes 
\begin{eqnarray}
    && J_{\perp(\bar n)} \left ( \bar n \cdot p', \omega_i, \mu, \nu \right )  
\nonumber \\
    &&= 1 + \frac{ \alpha_s (\mu) }{4 \pi} \frac{4}{3} \left [ \left ( \ln \frac{\mu^2}{n \cdot p' (\omega- \bar{n} \cdot p')}  - 2 \ln  \frac{\omega- \bar{n} \cdot p'}{ \omega_2- \bar{n} \cdot p'} - {1 \over 2} \right ) \ln  \frac{\mu^2}{n \cdot p' (\omega- \bar{n} \cdot p')}  - \ln  \frac{\nu}{\mu}  \right . \notag\\
	&& \quad \left . - \ln^2  \frac{\omega- \bar{n} \cdot p'}{ \omega_2- \bar{n} \cdot p'}  + 2 \ln  \frac{\omega- \bar{n} \cdot p'}{ \omega_2- \bar{n} \cdot p'}  \left ( \frac{\omega_2- \bar{n} \cdot p'}{ \omega_1} -\frac{3}{4} \right )  -\frac{\pi^2}{6} -\frac{1}{2} \right ] \, .
\end{eqnarray}

\subsection{The LCSR of $\Lambda_b \to p $ form factors at $\mathcal{O}(\alpha_s)$}
 Finally we present NLL resummmation improved factorized formulae for the invariant
 amplitudes entering the Lorenz decomposition of the correlation functions $\Pi_{\mu,a} (P,q)$ , 
 \begin{eqnarray}
   \Pi_{\perp(\bar n),V(A)} &=& f^{(2)}_{\Lambda_b}(\mu) \left [ U_1 \left ( \bar{n} \cdot p' ,\mu_h , \mu \right )  C_{\perp(\bar n),V(A)} (n \cdot p' ,\mu_h ) \right ] 
\nonumber \\
    && \int_{0}^{\infty} d\omega_1 \, d\omega_2 \frac{ 1 }{\omega_1 + \omega_2 - \bar{n} \cdot p' - i0 } \,  J_{\perp(\bar n)} \left ( \bar n \cdot p', \omega_i, \mu, \nu \right ) \, \phi_4(\omega_1,\omega_2,\mu) \, ,
\nonumber \\
   \Pi_{{n},V(A)} &=&  f^{(2)}_{\Lambda_b}(\mu) C_{{n} ,V(A)} (n \cdot p' ,\mu ) \int_{0}^{\infty} d\omega_1 \, d\omega_2 \frac{ 1}{\omega_1 + \omega_2 - \bar{n} \cdot p' - i0 }  \phi_4(\omega_1,\omega_2, \mu) \, .
\end{eqnarray}

Then we transform the factorized correlation functions into their dispersion forms with the help of the relations in \cite{wang_perturbative_2016} and apply the continuum subtraction and the Borel transformation to construct QCD sum rules. 
The form factors become,
\begin{eqnarray}\label{fTfinal}
    &&f_N(\nu) (n \cdot p' ) \, e^{-{m^2_p \over n \cdot p' \omega_M}} \left\{f^{T}_{\Lambda_b \to p} (q^2) , g^{T}_{\Lambda_b \to p} (q^2)\right\}  \notag\\
	&&= f^{(2)}_{\Lambda_b}(\mu) [U_1 (\bar{n} \cdot p' ,\mu_h , \mu )  C_{\perp ,V(A)} (n \cdot p' ,\mu_h ) ] \int_{0}^{\omega_s} d\omega  e^{-{\omega \over \omega_M}} {\psi}_{4,{\rm eff}}(\omega,\mu,\nu) ,
\end{eqnarray}
\begin{eqnarray}
   && f_N(\nu) (n \cdot p' ) \, e^{-{m^2_p \over n \cdot p' \omega_M}} \left\{f^{0}_{\Lambda_b \to p} (q^2) , g^{0}_{\Lambda_b \to p} (q^2)\right\}  \notag\\
   &&= f^{(2)}_{\Lambda_b}(\mu) [U_1 (\bar{n} \cdot p' ,\mu_h , \mu )  C_{\bar{n} ,V(A)} (n \cdot p' ,\mu_h ) ] \int_{0}^{\omega_s} d\omega   e^{-{\omega \over \omega_M}} {\psi}_{4,{\rm eff}}(\omega,\mu,\nu)  \notag\\
	&& \quad + f^{(2)}_{\Lambda_b}(\mu) \left(1-\frac{n \cdot p'}{m_{\Lambda_b}}\right) C_{{n} ,V(A)} (n \cdot p' ,\mu) \int_{0}^{\omega_s} d\omega   e^{-{\omega \over \omega_M}} \tilde{\psi}_{4}(\omega,\mu)  ,
\end{eqnarray}
\begin{eqnarray}
    &&f_N(\nu) (n \cdot p' ) \, e^{-{m^2_p \over n \cdot p' \omega_M}} \left\{f^{+}_{\Lambda_b \to p} (q^2) , g^{+}_{\Lambda_b \to p} (q^2)\right\}\notag\\
	&& = f^{(2)}_{\Lambda_b}(\mu) [U_1 (\bar{n} \cdot p',\mu_h , \mu )  C_{\bar{n} ,V(A)} (n \cdot p' ,\mu_h ) ] \int_{0}^{\omega_s} d\omega  e^{-{\omega \over \omega_M}} {\psi}_{4,{\rm eff}}(\omega,\mu,\nu) \notag\\
	&& \quad - f^{(2)}_{\Lambda_b}(\mu) \left(1-\frac{n \cdot p'}{m_{\Lambda_b}}\right) C_{{n} ,V(A)} (n \cdot p' ,\mu ) \int_{0}^{\omega_s} d\omega  e^{-{\omega \over \omega_M}} \tilde{\psi}_{4}(\omega,\mu)  ,
\end{eqnarray}
where we need to drop out $\mathcal{O}(\alpha_s^2)$ terms beyond the NLL approximation to conduct the numerical calculation, and 
\begin{eqnarray}
    {\psi}_{4,{\rm eff}}(\omega,\mu,\nu) &=&\tilde{\psi}_{4}(\omega,\mu) \left [ 1 +  \frac{\alpha_s(\mu)}{3 \pi}  \left ( \ln^2 \frac{\mu^2}{n \cdot p' \omega} - {5 \over 2} \ln {\mu^2 \over n \cdot p' \, \omega} - \ln {\nu \over \mu}  +\frac{\pi^2}{6} - 6 \right )  \right ] 
\nonumber \\
    && - \frac{2 \alpha_s(\mu)}{3 \pi} \left \{ \int_{\omega}^{\infty} {d \omega'}  \ln \frac{\omega' - \omega}{\omega}  \left [ {\tilde \psi_4(\omega' ,\mu ) \over \omega'} - \left (  \ln \frac{\mu^2}{n \cdot p' \omega} + \frac{11}{4} \right ) \frac{d }{d \omega'} { \omega \tilde \psi_4 (\omega',\mu) \over \omega'} \right ] \right . \notag\\
    && - \int_{0}^{\omega} \frac{ d \omega'}{\omega' - \omega} \left ( \ln \frac{\mu^2}{n \cdot p' (\omega - \omega')} - \ln {\omega - \omega' \over \omega} - {5 \over 4} \right ) \left [ \tilde{\psi}_4 (\omega',\mu) - \tilde{\psi}_4 (\omega,\mu) \right ] \notag\\
    && \left . - \int_{0}^{\omega} {d \omega' \over \omega'}  \ln \frac{\omega - \omega'}{\omega} \tilde \psi_4 (\omega',\mu) \right \} \, .
\end{eqnarray}
Because $\phi_4(u\omega_1,(1-u)\omega,\mu) $ is supposed to be independent on the momentum fraction $u$ as mentioned in \cite{wang_perturbative_2016}, 
for brevity, we define $\tilde{\psi}_4 (\omega, \mu)= \omega \phi_4(u\omega,(1-u)\omega,\mu)$. 

After the continuum subtraction, the integration bounds of $\omega'$ are supposed to be $\omega_s \sim \omega' \sim \lambda^2 m_b $. To avoid large logarithms, the factorization scale $\mu$ should be $\mu \sim \lambda m_b \sim m_p$.

\section{Numerical results}\label{sect4}
\subsection{Theory input parameters}
Here we  consider the exponential model $ \psi_4^{{\rm I}} (\omega , \mu) $ \cite{ball_distribution_2008,bell_light-cone_2013} as the parameterizations of the $\Lambda_b $-baryon LCDA $\psi_4 (\omega , \mu)$ at a soft scale. There are still two alternative parameterizations $\psi_4^{{\rm II}} (\omega,\mu), \psi_4^{{\rm III}} (\omega,\mu)$ \cite{ball_distribution_2008,bell_light-cone_2013}, which we use them to conduct the analysis of the model dependence.
\begin{align}
    \psi_4^{{\rm I}} (\omega,\mu) = & \frac{1}{\omega_0^2} e^{-{\omega \over \omega_0}} ,\notag\\
    \psi_4^{{\rm II}} (\omega,\mu) = & \frac{1}{\omega_0^2}  e^{-\left (\omega \over \omega_1 \right )^2} , & \omega_1 =& \sqrt{2} \omega_0 , \notag\\
    \psi_4^{{\rm III}} (\omega,\mu) = & \frac{1}{\omega_0^2} \left[1- \sqrt{(2- \frac{\omega}{\omega_2}) \frac{\omega }{\omega_2}} \right]  \theta (\omega_2-\omega )  , & \omega_2 = & \sqrt{ \frac{12}{10-3 \pi }} \omega_0.
\end{align}

 We remark that these models do not develop the radiative tail at large $\omega$ due to perturbative
 corrections, and they should be merely treated as an effective description of $\psi_4(\omega,\mu)$ at small $\omega$,
 where QCD factorization of the correlation functions is established.

The coupling $f^{(2)}_{\Lambda_b} (\mu) $ is taken from the NLO HQET sum rule calculation \cite{ball_distribution_2008}  ,
 \begin{eqnarray}
    f^{(2)}_{\Lambda_b} ({\rm 1 GeV}) =(3.0 \pm 0.5) \times 10^{-2} {\rm GeV^3} \, .
 \end{eqnarray}

 The normalization parameter $f_N(\nu)$ is determined in Lattice QCD \cite{PhysRevD.111.094517} ,
 \begin{eqnarray}
    f_N({\rm 1 GeV }) = (3.29 \pm 0.28) \times 10^{-3} {\rm GeV^2}.
 \end{eqnarray}

 For the choices of the renormalization and the factorization scales entering the NLL sum rules, the renormalization scale of the baryonic current $\nu$ and the
 factorization scale $\mu$ will be varied in the interval $1 {\rm GeV} \le \mu , \nu \le 2 {\rm GeV}$ around the default value $\mu = \nu = 1.5 {\rm GeV}$. The hard scale $\mu_h$ in the hard matching coefficients will be taken as $\mu_h = m_b $  with the variation in the range $[m_b /2 ,3 m_b/2]$. In addition, we adopt the $\rm \bar{MS}$ bottom-quark mass $\overline{m_b}(\overline{m_b})  = 4.193^{+0.022}_{-0.035} {\rm GeV}  $ \cite{ParticleDataGroup:2024cfk}.  

We choose the Borel parameter $\omega_M$ and the effective duality threshold $\omega_s$  for the $\psi_4^{{I}} $ as 
\begin{align}\label{omphi1}
   M^2 \equiv & n \cdot p' \omega_M = (1.5 \pm 0.5 ) {\rm GeV^2}  \,, & s_0 \equiv & n \cdot p' \omega_s = (2.25\pm 0.50 ) {\rm GeV^2} .
\end{align}
If we assume that the three models cause the same value $f^{T}_{\Lambda_b \to p } (0) $, keeping the Borel parameter and the effective duality threshold in \Eq{omphi1}, we may designate  the value of $\omega_0$ for all the models as follows, 
\begin{eqnarray}\label{omphi2}
	    &&\omega_0 = (450 \pm 150) {\rm MeV } \, (\rm Model \,\, I ) \, ,\notag\\
        &&\omega_0 = (599 \pm 152) {\rm MeV } \,  (\rm Model \,\, II ) \, ,\notag\\
          &&  \omega_0 = (500 \pm 132) {\rm MeV } \,  (\rm Model \,\, III ) \, .
\end{eqnarray}
 The parameterization of $\psi_4(\omega, \mu)$ governs its small-$\omega$ behavior, which is effectively controlled by the parameter $\omega_0$. Consequently, the resulting form factors exhibit a clear dependence on the chosen value of $\omega_0$. 
 To obtain form factors to a satisfactory approximation, it is therefore essential to select a reliable $\omega_0$. In this work, we determine $\omega_0$ for the three considered models following the strategy outlined in \cite{wang_ensuremathlambda_bensuremathrightarrowp_2009, huang_lambda_2023}.

 The effective threshold parameter $s_0$  should minimize the contributions of the unwanted hadronic states to reduce the systematic uncertainty induced by the parton-hadron duality approximation. 
 As for the heavy-light systems, the standard value of the threshold in the $X$ channel would be  $s_0^X = (m_X + \Delta_X)^2 $ and $\Delta_X$ is about $0.5 \, \rm GeV$ \cite{wang_ensuremathlambda_bensuremathrightarrowp_2009} . 
 So it is a good choice to designate $s_0 = 2.25 \rm GeV^2$.

For reducing the potential contributions of poorly known higher twist LCDAs of $\Lambda_b$-baryon, the sum rule predictions should be relatively stable in the allowed region for Borel mass $\omega_M$. 
We can impose the following conditionon the form factors \cite{wang_perturbative_2016},
\begin{eqnarray}
	\frac{\omega_M}{F^i_{\Lambda_b \to p}}\frac{\partial F^i_{\Lambda_b \to p}}{\partial \omega_M} \le 40\%,
\end{eqnarray}
where $F^i_{\Lambda_b \to p}$ stands for a general $\Lambda_b \to p $ form factor.  We will show that the form factors are rather stable in the selected region of $M^2$.

\subsection{Predictions for the $\Lambda_b \to p$ form factors}

\begin{figure}[htbp]
			\centering
		
			\subfigure[]
			{
				\begin{minipage}[b]{.3\linewidth}
					\centering
					\includegraphics[scale=0.5]{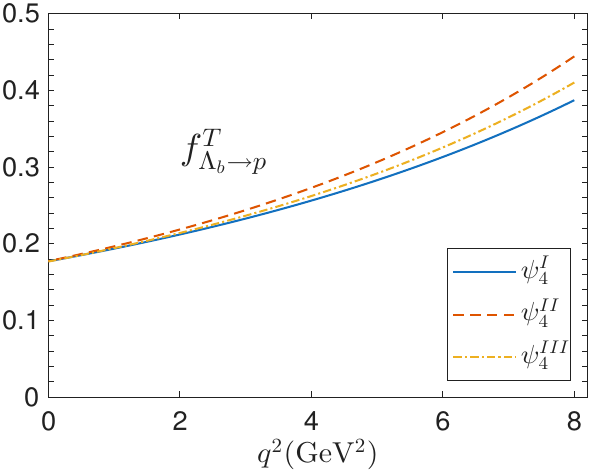}
				\end{minipage}
			}
			\subfigure[]
			{
				\begin{minipage}[b]{.3\linewidth}
					\centering
					\includegraphics[scale=0.5]{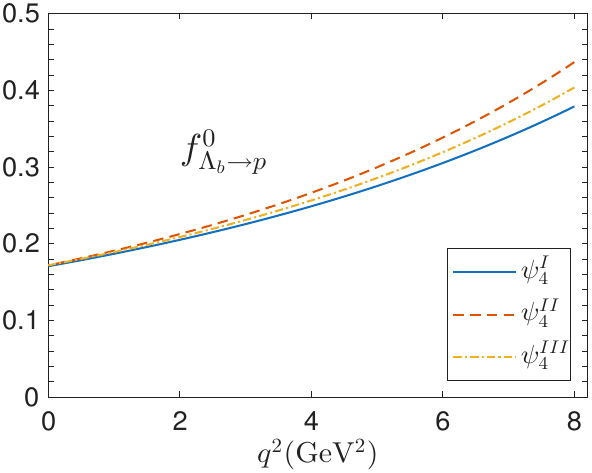}
				\end{minipage}
			}
			\subfigure[]
			{
				\begin{minipage}[b]{.3\linewidth}
					\centering
					\includegraphics[scale=0.5]{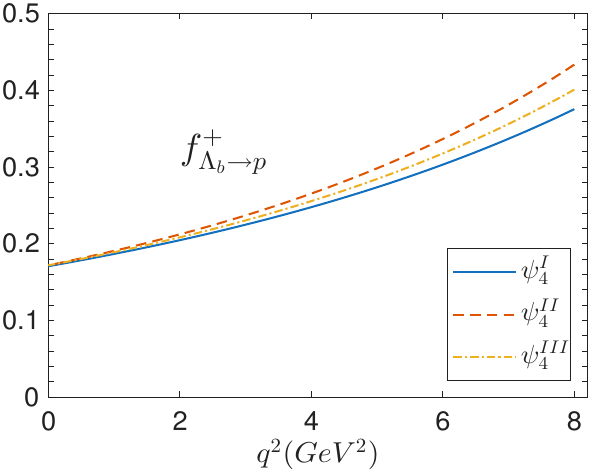}
				\end{minipage}
			} 
		
	\caption{The momentum-transfer dependence of the $\Lambda_b \to p $ form factors compute from LCSR with the fitted values of $\omega_0$ parameters presented in \Eq{omphi1}and \Eq{omphi2} for three different models of $\psi_4(\omega,\mu)$. Solid , dashed and dot dashed curves correspond to the sum rule predictions with the $\Lambda_b $-baryon LCDA $\psi^{\rm I}_4(\omega,\mu)$,$\psi^{\rm II}_4(\omega,\mu)$ and $\psi^{\rm III}_4(\omega,\mu)$}
	\label{ftf0f}	\end{figure}

 Our main purpose is to predict the momentum-transfer dependence of all the six $\Lambda_b \to p$ form factors assumed the reduced model dependence of $\psi_4 (\omega,\mu)$. 
After determining the  necessary input parameters and analysing the form factors from the sum rules numerically, we can indicate the expected insensitivity of the parameterizations of $\psi_4(\omega,\mu )$ in Fig.\ref{ftf0f}. $\,$ 
 In the low-$q^2$ region, the predictions from various models are in excellent agreement. A noticeable deviation emerges around $q^2 = 8~\text{GeV}^2$. However, this does not affect the validity of our analysis since the sum rule derivation is strictly justified only at low $q^2$, consistent with the region of agreement.

\begin{figure}[htbp]
			\centering
		
			\subfigure[]
			{
				\begin{minipage}[b]{.3\linewidth}
					\centering
					\includegraphics[scale=0.5]{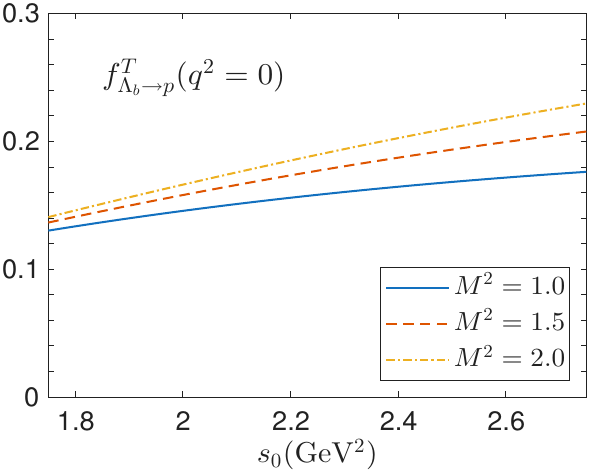}
				\end{minipage}
			}
			\subfigure[]
			{
				\begin{minipage}[b]{.3\linewidth}
					\centering
					\includegraphics[scale=0.5]{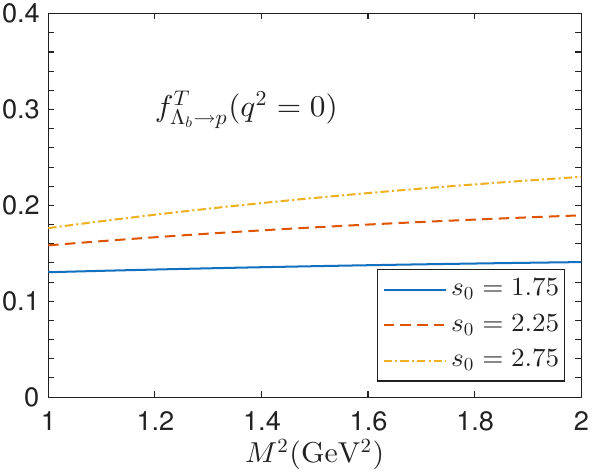}
				\end{minipage}
			}
			\subfigure[]
			{
				\begin{minipage}[b]{.3\linewidth}
					\centering
					\includegraphics[scale=0.5]{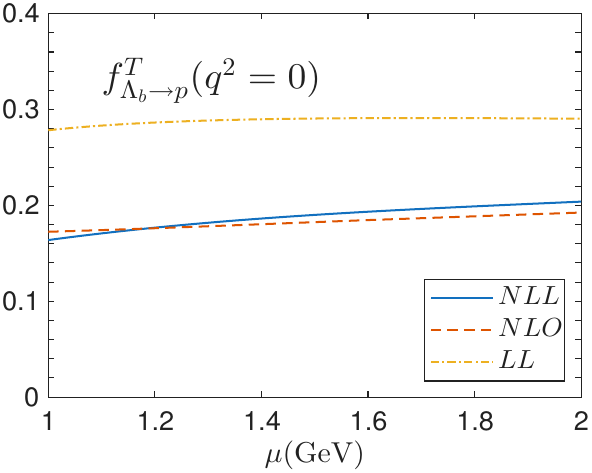}
				\end{minipage}
			} 
		
	\caption{Dependence of  $f^T_{\Lambda_b \to p} (0) $ on the threshold parameter (left), on the Borel parameter(middle), and on the factorization scale (right).    Solid , dashed and dot dashed curves correspond to the sum rule predictions with $M^2=1.5 \rm GeV^2 $, $M^2=2 \rm GeV^2 $, $M^2=2.5\rm  GeV^2 $ (left) and $s_0=1.42 \rm GeV^2 $, $s_0=1.52 \rm GeV^2 $, $s_0=1.62 \rm GeV^2 $(middle). The label "LL" , "NLO" and "NLL" (right) represent the sum rule predictions at LL, NLO and NLL accuracy. All the other input parameters are fixed at their central values with the $\Lambda_b$-baryon LCDA $\phi^I_4 (\omega,\mu)$.}
	\label{m2s0}	\end{figure}

In Fig. \ref{m2s0}, the sum-rule results for $f^T_{\Lambda_b \to p}(0)$  have weak $M^2$-dependence, meeting the stability criterion , yet their strong $s_0$-dependence implies a inherent systematic error that limits the ultimate precision.
Furthermore, both the leading-logarithmic (LL) and NLL improved sum rules show little sensitivity to the factorization scale $\mu$ across its allowed range. This behavior is consistent with the expected factorization-scale independence of the physical form factors.  
In addition, compared to the pure one-loop fixed-order correction, the resummation of parametrically large logarithms in the hard matching coefficient has only a minor impact on the sum-rule predictions for $f^T_{\Lambda_b \to p}(0)$.

\begin{figure}[htbp]
			\centering

			\subfigure[]
			{
				\begin{minipage}[b]{.45\linewidth}
					\centering
					\includegraphics[scale=0.5]{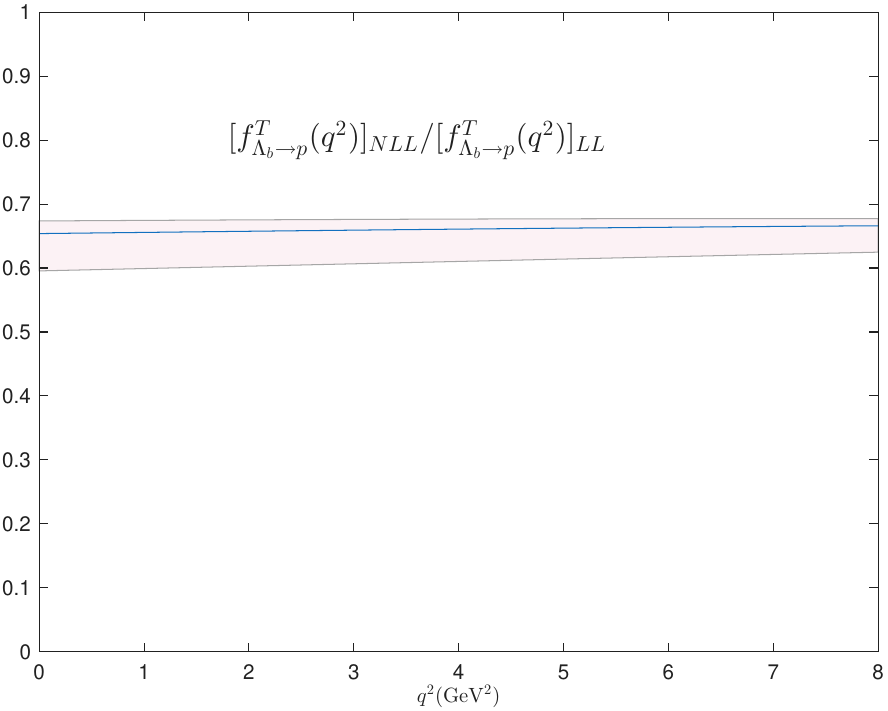}
				\end{minipage}
			}
	\subfigure[]
			{
				\begin{minipage}[b]{.45\linewidth}
					\centering
					\includegraphics[scale=0.5]{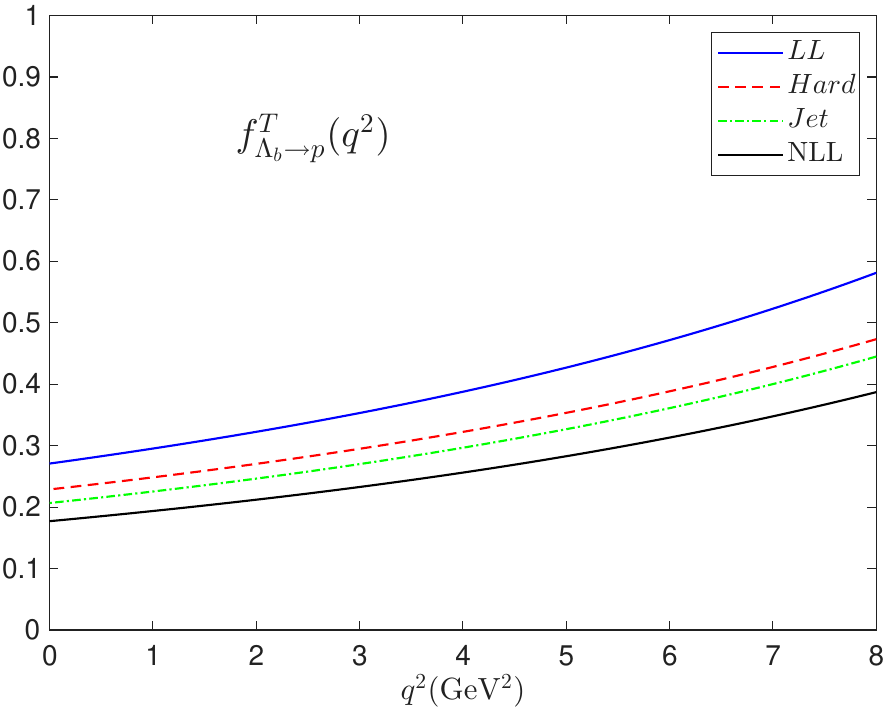}
				\end{minipage}
			}
		
	\caption{Compared with the predictions at LL and NLL accuracy, the contribution to the sum rules of  $f^T_{\Lambda_b \to p} (q^2) $ from the NLO hard and the NLO jet function (left panel)  and the momentum transfer dependence of the ratio $[f^T_{\Lambda_b \to p} (q^2)]_{NLL} / [f^T_{\Lambda_b \to p} (q^2)]_{LL} $ with theory uncertainty from varying the renormalization and the factorization scales(right panel).}
		\label{Nllll}\end{figure}

More importantly, Fig.\ref{Nllll} (left panel) indicates that the perturbative $\mathcal{O}(\alpha_s)$ correction reduces the tree-level sum rule prediction  to approximately $65\%$ of their original value, 
which implys the importance of QCD radiative effect in baryonic sum  rule applications. 
Then we consider the hard and the hard-collinear corrections of $f^T_{\Lambda_b \to p} (q^2)$ in Fig.\ref{Nllll} (right panel), 
which are defined as replacing $\psi_{4,eff}(\omega',\mu,\nu) $ in \Eq{fTfinal} by $\tilde{\psi_4}(\omega')$ for the former and as defined as replacing $U_1 (\bar{n} \cdot p'/2 ,\mu_h , \mu )  C_{\perp(\bar{n}) ,V(A)} (n \cdot p' ,\mu_h ) $ by one for the latter.  
It is obvious that the NLO jet  function is the dominant $\alpha_s$ correction at one loop, which highlights the importance of the perturbative matching calculation at the hard-collinear scale.

For the proton energy dependence of the form factor $f^T_{\Lambda_b \to p} (q^2)$ from the sum rules at LL and NLL accuracy, we introduce the following ratio originally proposed in  \cite{fazio_scet_2008}
\begin{eqnarray}
	R_1(E_p) = \frac{f^T_{\Lambda_b \to p} (n \cdot p')}{f^T_{\Lambda_b \to p} (m_{\Lambda_b})},
\end{eqnarray}
\begin{figure}[htbp]
			\centering
					\includegraphics[scale=0.5]{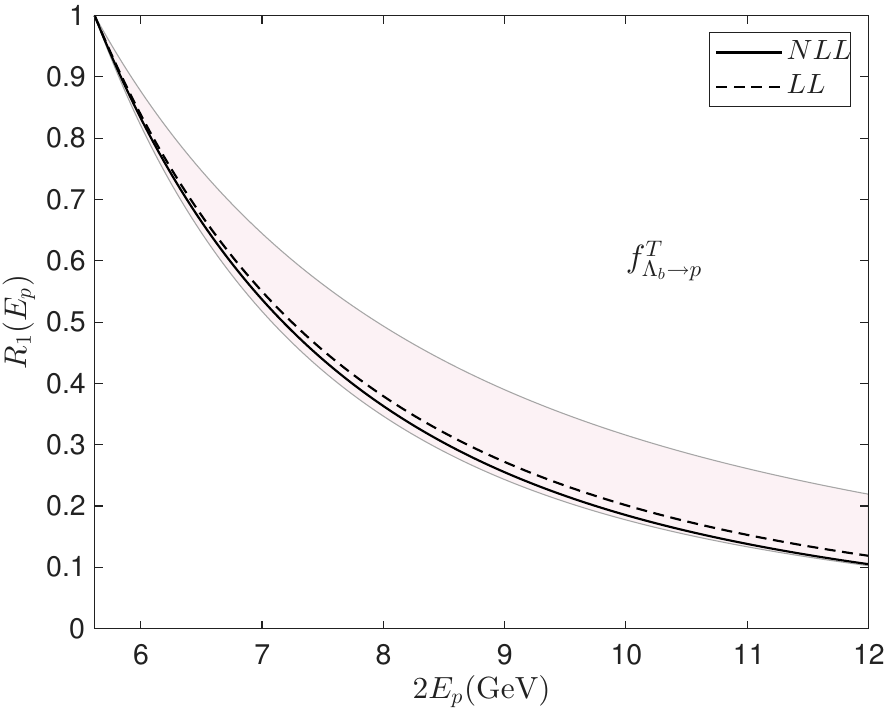}
	\caption{Dependence of the ratio $R_1(E_p)$ on the proton energy $E_p$. The dashed and solid lines are obtained from the LL and NLL sum rule predictions, respectively. The two red curves refer to a pure $1/ E^2_p$  and a pure $1/E_p^3$ dependence.} 
\label{R1}	
\end{figure}
where We employ the proton energy $E_p$ as the argument for the form factors, replacing $q^2$. Here, $E_p$ is equivalent to $\frac{1}{2} n \cdot p'$.  As shown in Fig. \ref{R1}, both the LL and NLL sum rule predictions favors evidently a $1/E_p^3$ behaviour in consistent with the power counting analysis. 

While the explicit analysis is performed for $f^T_{\Lambda_b \to p}$, the resulting conclusions can be extended to all form factors by the same reason.

Because the light-cone operator-product expansion of the correction function $\Pi_{\mu,a } (P,q)$ cna only be justified at low $q^2$, we have to apply the simplified $z$-series parameterization \cite{PhysRevD.79.013008} to extrapolate the sum rule predictions for the $\Lambda_b \to p $ form factors at $q^2 \le q^2_{\rm max} = 8 GeV^2 $ towards large momentum transfer $q^2$. The standard transformation is 
\begin{eqnarray}
	z(q^2,t_0) = \frac{\sqrt{t_+ -q^2 } - \sqrt{t_+ -t_0 } }{\sqrt{t_+ -q^2 } +\sqrt{t_+ -t_0 } } \, ,
\end{eqnarray}
which transforms the cut $q^2$-plane onto the disk $|z(q^2,t_0)| \le 1$ on the complex $z$-plane.

Following  \cite{detmold_mathrmensuremathlambda_bensuremathrightarrowpensuremathellensuremath-overlineensuremathnu_ensuremathell_2015}, we take $t_+ = (m_B + m_{\pi})^2$ for all $\Lambda_b \to p $ form factors because the form factors can be analytical functions in the complex $q^2-$plane.    The auxiliary parameter $t_0 $ will be chosen as the maximum transfer momentum in the physical region $q^2_{\rm max} = (m_{\Lambda_b} - m_p)^2 $.  The helicity from factors are induced by the helicity-projected weak current operator with definite spin-parity quantum numbers \cite{wang_perturbative_2016}.

\begin{table}[http]
\centering

\begin{tabular}{cccccccc}
\hline\hline
  Form factor &$B(J^P)$ \cite{wang_perturbative_2016} & $m^f_{\rm pole} \, (\rm GeV)$ \cite{detmold_mathrmensuremathlambda_bensuremathrightarrowpensuremathellensuremath-overlineensuremathnu_ensuremathell_2015} & $f (0)$ & $a^f_0 $& $a^f_1 $ & $\rho_{01}$\\
\hline
$f^{T}_{\Lambda_b \to p } (q^2) $ & $B^* (1^-)$ & 5.33 &$0.177\pm 0.112$ &$0.61 \pm  0.41$  & $-1.30 \pm 1.11 $&-0.985\\
$f^{0}_{\Lambda_b \to p } (q^2) $ & $B (0^+) $ &5.66&$0.171\pm 0.108$ &$0.64 \pm  0.42$  & $-1.42 \pm 1.12 $ &-0.988\\
$f^{+}_{\Lambda_b \to p } (q^2) $ & $B^* (1^-)$ & 5.33&$0.171 \pm 0.108$&$0.59 \pm 0.40 $  & $-1.27 \pm 1.07 $ &-0.986\\
$g^{T}_{\Lambda_b \to p } (q^2) $ & $B_{1} (1^+) $ &5.71&$0.177\pm 0.112 $ &$0.65 \pm  0.43$  & $-1.44 \pm 1.15 $& -0.987\\
$g^{0}_{\Lambda_b \to p } (q^2) $ & $B (0^-) $ &5.28&$0.171 \pm 0.108$ &$0.60 \pm  0.40$  & $-1.28\pm 1.08 $& -0.986\\
$g^{+}_{\Lambda_b \to p } (q^2) $ & $B_{1} (1^+) $ &5.71&$0.171 \pm 0.108$&$0.64 \pm 0.41 $  & $-1.41 \pm 1.12 $& -0.987\\
\hline\hline

\end{tabular}

\caption{Summary of the masses of low-lying resonances produced by the helicity-projected weak current operators $\bar{u} \Gamma_{\mu,a} b $ in QCD, the central value of all the form factors $f(0)$  and the shape parameters $a^f_i$ by matching the $z$-series parameterizations to our NLL sum rule predictions. }
\label{masspole}
\end{table}

\begin{table}[http]
\centering

\begin{tabular}{cccccccccc}
\hline\hline
  Form factor & $a^f_0 $& $a^f_1 $&$a^f_2 $ &  $\rho_{01}$ & $\rho_{02}$ &$\rho_{12}$\\
\hline
$f^{T}_{\Lambda_b \to p } (q^2) $ &$0.491 \pm  0.059$  & $-0.21 \pm 0.69 $&$-2.10 \pm1.50  $ &-1.000&0.999&-1.000 \\
$f^{0}_{\Lambda_b \to p } (q^2) $ & $0.359 \pm  0.042$  & $-0.11 \pm 0.50 $&$-1.56 \pm 1.14 $&-1.000&0.999&-1.000 \\
$f^{+}_{\Lambda_b \to p } (q^2) $ &$0.384 \pm 0.046 $  & $0.03 \pm 0.54 $&$-2.03 \pm 1.21$&-1.000&0.999&-1.000 \\
$g^{T}_{\Lambda_b \to p } (q^2) $  &$0.354 \pm  0.037$  & $-0.54 \pm 0.46 $&$-0.36\pm1.07 $&-1.000&0.999&-1.000\\
$g^{0}_{\Lambda_b \to p } (q^2) $ &$0.403 \pm  0.047$  & $-0.20\pm 0.55 $&$-1.57 \pm 1.23 $ &-1.000&0.999&-1.000\\
$g^{+}_{\Lambda_b \to p } (q^2) $ & $0.340 \pm 0.035 $  & $-0.25 \pm 0.43 $&$-1.03\pm 1.00$&-1.000&0.999&-1.000\\
\hline\hline

\end{tabular}

\caption{Summary of the shape parameters $a^f_i$ for all the form factors through the combined fitting between our NLL sum rule predictions and the lattice results. }
\label{masspoletotal}
\end{table}

Keeping the series expansion of the form factors to the second power of $z$-parameter, the parameterizations is 
\begin{eqnarray}
	f_{\Lambda_b \to p} (q^2) = \frac{1}{1-q^2/m^f_{\rm pole} } \left\{ a^f_0+ a^f_1 z(q^2,t_0) + a^f_2 z^2 (q^2,t_0) \right\},
\end{eqnarray}
where $	f_{\Lambda_b \to p} (q^2) $ represents the all six $\Lambda_b \to p$ form factors and the corresponding pole masses $m^f_{\rm pole}$ are set to the values given in Table \ref{masspole}. 
The shape parameters $a^f_i$ can be determined by matching the $z$-series parameterizations to the predictions of the form factors. $\rho_{ij}$ is the correlation coefficient between $a^f_i$ and $a^f_j$.

For our NLL sum-rule predictions in the large-recoil region ($0 \le q^2 \le q^2_{\mathrm{max}} = 8~\mathrm{GeV}^2$), we observe that a first-order expansion in the $z$-parameter fits the form factors  as well as a second-order expansion. For simplicity, we adopt the first-order $z$-expansion to perform a correlated $\chi^2$ fit , separately for each form factor .
To this end, we first generate three correlated LCSR data points at $q^2 = \left\{0, 2 , 4\right\}~\mathrm{GeV}^2$ , using $N=500$ ensembles of the input parameter set (which includes $M^2$, $s_0$, $\mu$, $\nu$, $\mu_h$ and $\omega_0$). The parameters in each ensemble are randomly sampled to construct the covariance matrix for this combined dataset. 
Since the data points are nearly linearly correlated, in order to incorporate more data points into the fitting process, we assume the presence of an additional small systematic error that reduces the off-diagonal elements of the covariance matrix to 95\% of their original values.
The corresponding loss function is built as a correlated $\chi^2$ statistic,
\begin{eqnarray}
	\chi^2 = \left\{ \vec{ F }_{\rm model } - \vec{ F }_{\rm data }  \right\}^{T} \cdot C_{\rm cov}^{-1} \cdot \left\{ \vec{ F }_{\rm model } - \vec{ F }_{\rm data }  \right\} \, ,
\end{eqnarray}
where $\vec{ F }_{\rm data }$ denotes the stacked vector of all input form-factor values, $\vec{F}_{\text{model}}$ is the corresponding vector of parameterized values, and ${C}_{\rm cov}$ is the associated total covariance matrix.
The $\chi^2$ is minimized to obtain the final fitted parameters.
Although only two data points are used, the fitted result agrees well with our original sum-rule predictions because the fit properly accounts for the correlation between them. The resulting fit parameters are listed in Table \ref{masspole}.


There is another widely used parameterization of the $\Lambda_b \to p$ form factors, $V-A$ form factor,
\begin{eqnarray}\label{hamV}
    &&\bra{{p} (p' ,s'  )   } \bar{u} \gamma_{\mu} b \ket{\Lambda_b (P,s )} =\notag\\
	&& u_p((p',s' ))  \biggl\{ F_1(q^2) \gamma_{\mu} + F_2(q^2) i\sigma_{\mu \nu} \frac{q^{\nu}}{m_{\Lambda_b}} + F_3(q^2) \frac{q_{\mu}}{m_{\Lambda_b}}    \biggr\}   u_{\Lambda_b} (P,s),
\end{eqnarray}
\begin{eqnarray}\label{hamA}
   && \bra{{p} (p' ,s'  )   } \bar{u} \gamma_{\mu}  \gamma_5 b \ket{\Lambda_b (P,s )} =\notag\\
   && u_p ((p',s' ))  \biggl\{   G_1(q^2) \gamma_{\mu} + G_2(q^2) i\sigma_{\mu \nu} \frac{q^{\nu}}{m_{\Lambda_b}} + G_3(q^2) \frac{q_{\mu}}{m_{\Lambda_b}}   \biggr\}    \gamma_5 u_{\Lambda_b} (P,s).
\end{eqnarray}
We list the $V-A$ form factor at $q^2=0$ in Table \ref{masspoleVA},  which is compared with the results from other approaches  .
We observe that the $F_1(0)$ and $G_1(0)$ we predict is consistent with the results from most other approaches within the 1-2 $\sigma$, despite a significant discrepancy with the results from \cite{huang_lambda_2023}.  
This difference can be understood by noting that while both studies use a similar LCSR framework, our calculation incorporates NLL resummation, whereas \cite{huang_lambda_2023} remains at leading order. As noted previously, the one-loop QCD radiative correction introduces a reduction of approximately $65\%$, which explains the observed shift and supports the validity of our result.

However, $F_2(0)$ and $G_2(0)$ we predict generally deviate by one order of magnitude from other results. This is a direct consequence of our leading-power  approximation. 
As also observed in \cite{huang_lambda_2023}, the LP current operator adopted here predominantly determines $F_1(0)$ and $G_1(0)$, while higher-power contributions mainly influence $F_2(0)$, $F_3(0)$, $G_2(0)$, and $G_3(0)$. Although the factorization is more straightforward at leading power, 
a future analysis incorporating higher power effects within the LCSR framework will be necessary to achieve a more complete and precise description of $\Lambda_b \to p $ form factors.

\begin{table}[http]
\centering

\begin{tabular}{ccccccc}
\hline\hline
  &  $F_1(0) $& $F_2(0) $&$G_1(0) $& $G_2(0) $&\\
\hline
This work & $0.171 \pm  0.076$ & $ -0.006 \pm 0.002 $ &$0.171 \pm  0.076$&$ 0.008 \pm 0.003$\\ 
heavy-LCSR-LP \cite{huang_lambda_2023} & $0.27 \pm 0.11$ & $-0.045 \pm 0.017$ & $0.27 \pm 0.11$& $-0.045\pm 0.017$\\
light-LCSR-$\mathcal{A}$ \cite{khodjamirian_form_2011} &$0.14^{+0.03}_{-0.03}$ & $-0.054^{+0.016}_{-0.013}$&$0.14^{+0.03}_{-0.03}$ & $-0.028^{+0.012}_{-0.009}$\\
light-LCSR-$\mathcal{P}$ \cite{khodjamirian_form_2011}&$0.12^{+0.03}_{-0.04}$ & $-0.047^{+0.015}_{-0.013}$&$0.12^{+0.03}_{-0.03}$ & $-0.016^{+0.007}_{-0.005}$\\
PQCD-Exponential \cite{han_lambda_2022} &$0.27 \pm 0.12$ & $0.008 \pm 0.005$ & $0.31 \pm 0.13$ & $0.014 \pm 0.010$\\
CCQM \cite{PhysRevD.90.114033}  &0.080 &-0.036&0.007&-0.001\\
RQM \cite{PhysRevD.94.073008}& 0.169&-0.050& 0.196&-0.0002\\
LFQM \cite{PhysRevD.80.094016} & 0.1131&-0.0356&0.1112&-0.0097\\
LQCD \cite{detmold_mathrmensuremathlambda_bensuremathrightarrowpensuremathellensuremath-overlineensuremathnu_ensuremathell_2015} &$0.22 \pm 0.08$ &$0.04 \pm 0.12$ &$0.12 \pm 0.14 $ &$0.04 \pm 0.31$\\
\hline\hline

\end{tabular}

\caption{Summary of  all the $V-A$ form factors from our NLL sum rule predictions and the comparison with the ones from other work. }
\label{masspoleVA}
\end{table}

\begin{figure}[htbp]
			\centering
		
			\subfigure[]
			{
				\begin{minipage}[b]{.3\linewidth}
					\centering
					\includegraphics[scale=0.35]{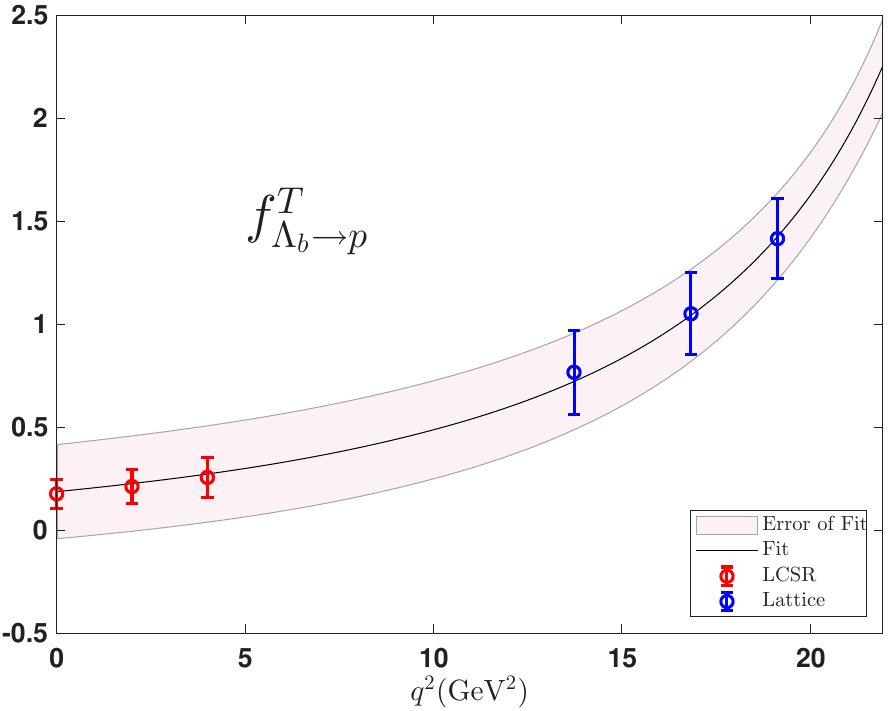}
				\end{minipage}
			}
			\subfigure[]
			{
				\begin{minipage}[b]{.3\linewidth}
					\centering
					\includegraphics[scale=0.35]{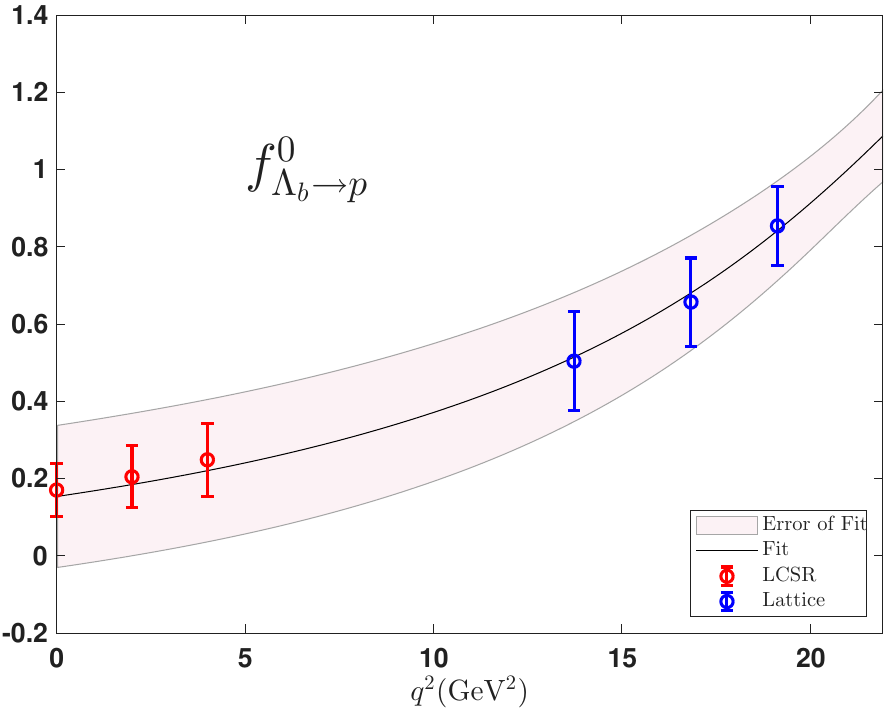}
				\end{minipage}
			}
			\subfigure[]
			{
				\begin{minipage}[b]{.3\linewidth}
					\centering
					\includegraphics[scale=0.35]{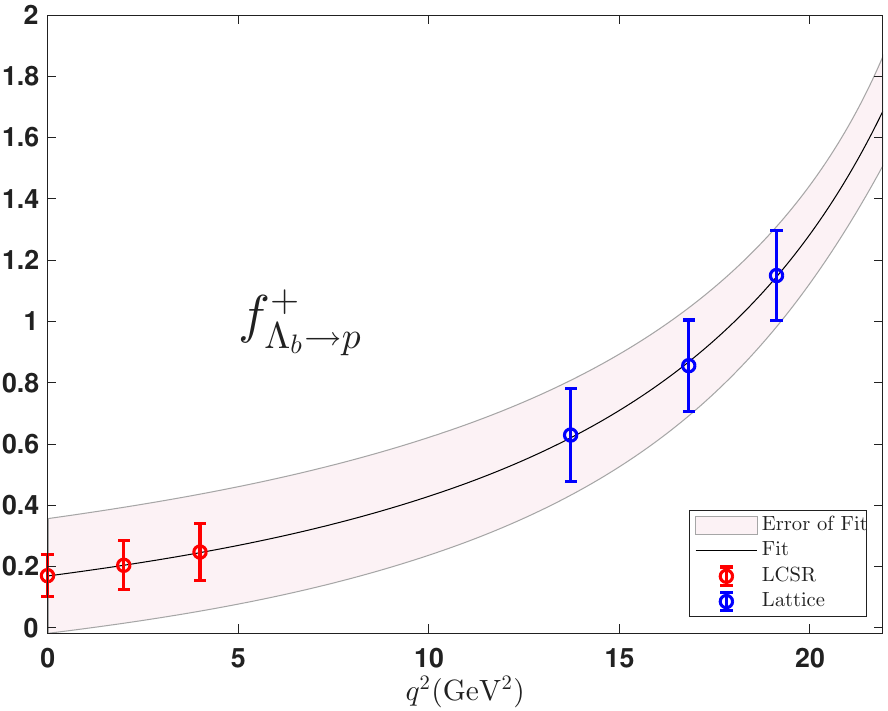}
				\end{minipage}
			} 
		
			\subfigure[]
			{
				\begin{minipage}[b]{.3\linewidth}
					\centering
					\includegraphics[scale=0.35]{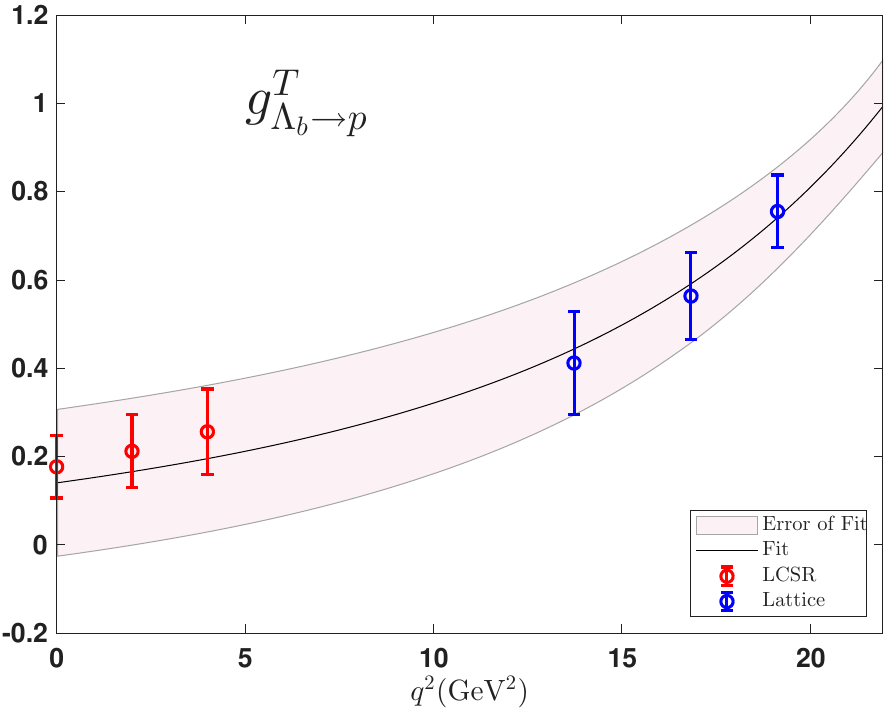}
				\end{minipage}
			}
			\subfigure[]
			{
				\begin{minipage}[b]{.3\linewidth}
					\centering
					\includegraphics[scale=0.35]{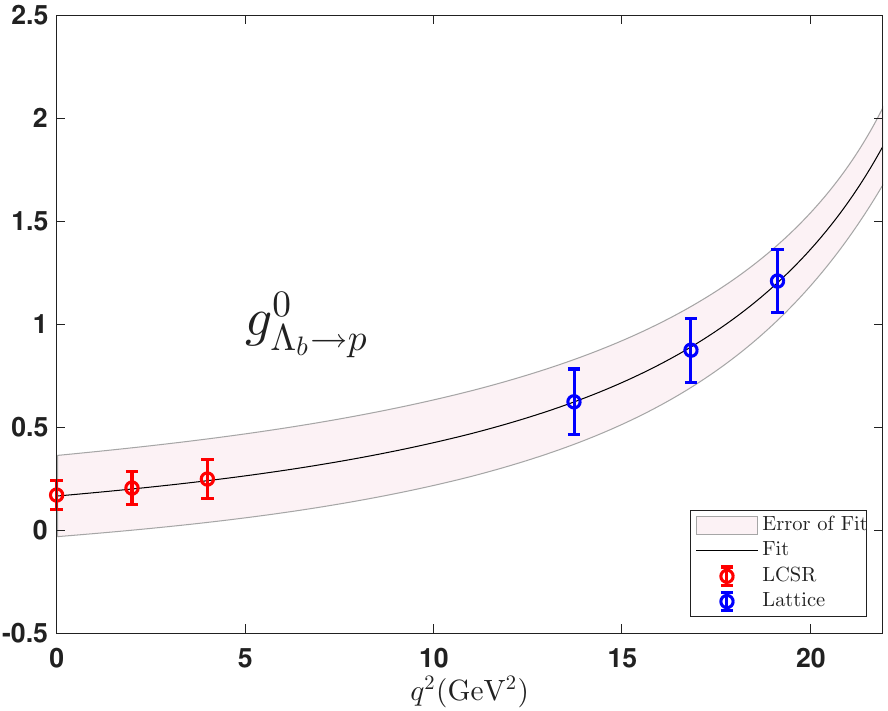}
				\end{minipage}
			}
			\subfigure[]
			{
				\begin{minipage}[b]{.3\linewidth}
					\centering
					\includegraphics[scale=0.35]{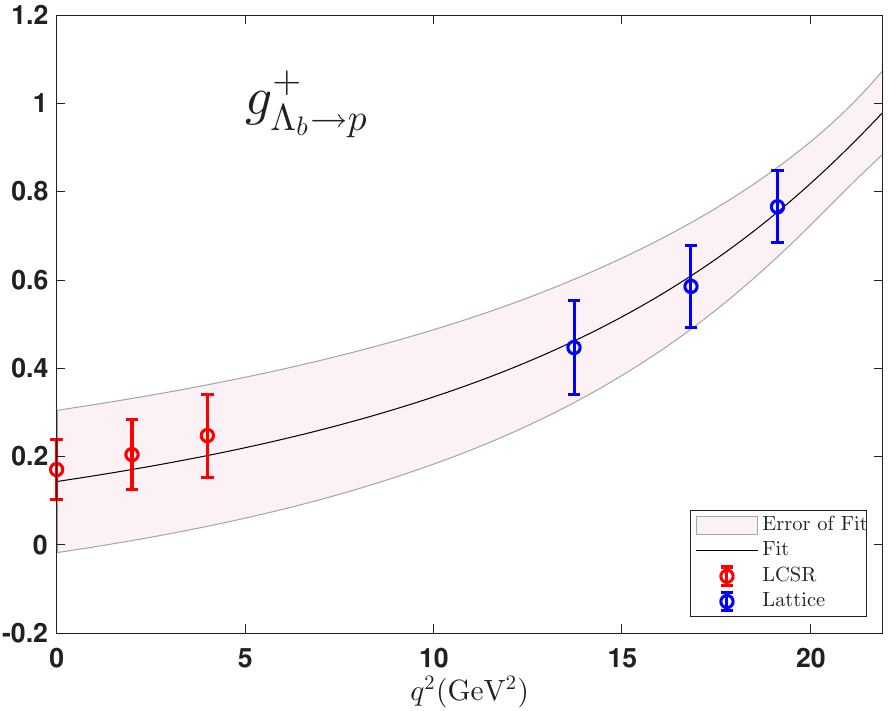}
				\end{minipage}
			} 
	\caption{ All the $\Lambda_b \to p $ form factors computed from the LCSR approach at NLL accuracy and fitted to the $z$-series parameterizations. The black solid  curve refers to the central values of the predictions from the combined fitting between the LCSR predictions and the lattice calculations. The red band represents the theory uncertainty of the fitting results. }
	\label{lka}	\end{figure}

We then perform a combined fit of our predictions and the lattice QCD results from \cite{detmold_mathrmensuremathlambda_bensuremathrightarrowpensuremathellensuremath-overlineensuremathnu_ensuremathell_2015} using a similar methodology.  The data points for the fit consist of our LCSR predictions at $q^2 = \left\{0, 2, 4 \right\}~\mathrm{GeV}^2$ and the lattice results at $q^2 = \left\{{13.7, 16.8 ,19.1 }\right\}~\mathrm{GeV}^2$.
The fit proceeds by randomly sampling the input parameter space.
After minimizing the $\chi^2$, the final fitted parameters are listed in Table \ref{masspoletotal}. The resulting $q^2$-dependence of the form factors is shown in Fig. \ref{lka}.
The fit result is in good agreement with both our LCSR predictions and the lattice QCD determinations.

\subsection{Phenomenological applications}

In the following, we aim at exploring the phenomenological applications of the obtained $\Lambda_b \to p$ form factors. These serve as fundamental ingredients for the theory description  of $\Lambda_b \to p \ell^- \bar{\nu}_{\ell} $ decays.
According to  \cite{PhysRevD.104.013005,zwicky_endpoint_2022,azizi_effects_2021}, we introduce helicity amplitudes, which are defined by
\begin{align}
	H^{V,A}_{\lambda_{p} ,\lambda_{W^-}} = & \varepsilon^{\dagger \mu}(\lambda_{W^-})  \bra{p(\lambda_p)} V(A)\ket{\Lambda_b (\lambda_{\Lambda_b})} \, ,
\end{align}
where $\lambda_{\Lambda_b},\lambda_{p},\lambda_{W^-}$ denote the helicity of the $\Lambda_b$ baryon, proton and off-shell $W^-$, respectively, which mediate semileptonic decays. The helicity vector $\varepsilon^{ \mu}(\lambda_{W^-})$ satisfies the following completeness relation 
\begin{align}
	\sum_{\lambda,\lambda' \in \left\{t,\pm,0\right\}}  \varepsilon^{* \mu}(\lambda)\varepsilon^{ \nu}(\lambda')  G_{\lambda \lambda'} = & g^{\mu\nu}  \,, &  G_{\lambda\lambda'} = & {\rm diag}(1,-1,-1,-1),
\end{align}  
where the first entry in $G_{\lambda\lambda'}$ refers to $\lambda=\lambda'=t$.

The helicity amplitudes $H^{V,A}_{\lambda_{p} ,\lambda_{W^-}}$ can be expressed as function of the form factors \cite{wang_perturbative_2016},
\begin{align}
	& H^{V}_{\frac{1}{2},0} = \frac{\sqrt{Q_-}}{\sqrt{q^2}} \left[M_+ F_1(q^2) - \frac{q^2}{m_{\Lambda}} F_2(q^2) \right] \,, && H^{V}_{\frac{1}{2},1} = {\sqrt{2Q_-}} \left[F_1(q^2) - \frac{M_+}{m_{\Lambda}} F_2(q^2) \right] \,,\,\notag\\
	& H^{V}_{\frac{1}{2},t} = \frac{\sqrt{Q_+}}{\sqrt{q^2}} \left[M_- F_1(q^2) +\frac{q^2}{m_{\Lambda}} F_3(q^2) \right] \,, && H^{A}_{\frac{1}{2},0} = \frac{\sqrt{Q_+}}{\sqrt{q^2}} \left[M_- G_1(q^2) +\frac{q^2}{m_{\Lambda}} G_2(q^2) \right] \,,\, \notag\\
	& H^{A}_{\frac{1}{2},1} = {\sqrt{2Q_+}} \left[G_1(q^2) + \frac{M_-}{m_{\Lambda}} G_2(q^2) \right] \,, && H^{A}_{\frac{1}{2},t} = \frac{\sqrt{Q_-}}{\sqrt{q^2}} \left[M_+ G_1(q^2) -\frac{q^2}{m_{\Lambda}} G_3(q^2) \right] \,,\, \notag\\
	& H^{V}_{- \lambda_{\Lambda_b}, - \lambda_{W^-} }  = H^{V}_{\lambda_{\Lambda_b},  \lambda_{W^-} }  \,, && H^{A}_{- \lambda_{\Lambda_b}, - \lambda_{W^-} }  = -H^{A}_{\lambda_{\Lambda_b},  \lambda_{W^-} } ,
\end{align}
where $Q_{\pm} =(m_{\Lambda_b }\pm m_p)^2 -q^2 $ and $M_{\pm}= m_{\Lambda_b} \pm m_p$. The total helicity amplitudes are then written as 
\begin{eqnarray}
	 H_{ \lambda_{\Lambda_b},  \lambda_{W^-} } = H^{V}_{\lambda_{\Lambda_b},  \lambda_{W^-} } -H^{A}_{\lambda_{\Lambda_b},  \lambda_{W^-} }.
\end{eqnarray}
So the differential angular distribution for the $\Lambda_b \to p \ell^- \bar{\nu}_{\ell}$ has the following form 
\begin{eqnarray}
	\frac{d\Gamma(\Lambda_b \to p \ell^- \bar{\nu}_{\ell})}{dq^2 d\cos \theta_{\ell}}  = \frac{G_F^2 |V_{ub}|^2 q^2 |\vec{p'}|}{512 \pi^3 m_{\Lambda_b}^2} \left(1- \frac{m_{\ell}^2}{q^2}\right)^2 \left(A_1 + \frac{m_{\ell}^2}{q^2} A_2 \right) \, , 
\end{eqnarray}
where $G_F$ is the Fermi constant, $V_{ub}$ is the CKM matrix element, $m_{\ell}$ is the lepton mass $(\ell= e,\mu,\tau)$, $\theta_{\ell}$ is the angle between the positive direction of the three-momentum of the final proton and the lepton in the $q^2$ rest frame, and   $\vec{p'}$ is the three-momentum of the proton,
\begin{align}
	|\vec{p'}|= & {1 \over 2 m_{\Lambda_b}} \sqrt{m_{\Lambda}^4 +m_p^4 + q^4 - 2 (m_{\Lambda_b}^2m_p^2 +m_p^2q^2  +m_{\Lambda_b}^2q^2)} \,,
\nonumber \\
	A_1 = & 2 \sin^2 \theta_{\ell} \left ( H^2_{\frac{1}{2},0}+H^2_{-{1 \over 2},0} \right ) + (1- \cos \theta_{\ell})^2 H^2_{\frac{1}{2},1} + (1+\cos \theta_{\ell})^2 H^2_{-{1 \over 2},-1}\,,\,
\nonumber \\
    A_2 =& 2 \cos^2 \theta_{\ell} \left (H^2_{\frac{1}{2},0}+H^2_{-{1 \over 2},0} \right )  +  \sin^2 \theta_{\ell} \left (H^2_{\frac{1}{2},1}+H^2_{-{1 \over 2},-1} \right )  +2 \left (H^2_{\frac{1}{2},t}+H^2_{-{1 \over 2},t} \right ) 
\nonumber \\
    &+4 \cos \theta_{\ell} \left (H_{\frac{1}{2},t}H_{\frac{1}{2},0}+H_{-{1 \over 2},t}H_{-{1 \over 2},0} \right ) \, .
\end{align}

The differential decay rate can be obtained by integrating with respect to $\cos \theta_{\ell}$ is 
\begin{eqnarray}
	\frac{d\Gamma(\Lambda_b \to p \ell^- \bar{\nu}_{\ell})}{dq^2 }  = \int_{-1}^{1}  \frac{d\Gamma(\Lambda_b \to p \ell^- \bar{\nu}_{\ell})}{dq^2 d \cos \theta_{\ell}}  d \cos \theta_{\ell}.
\end{eqnarray}
The leptonic forward-backward asymmetry $A_{FB}$, final state hadron polarization $P_B$  and the lepton polarization $P_l$ are defined as  \cite{huang_lambda_2023}
\begin{eqnarray}
    A_{FB}(q^2) &=& \frac{ \int_{0}^{1}  \frac{d \Gamma(\Lambda_b \to p \ell^- \bar{\nu}_{\ell})}{d q^2 d \cos \theta_{\ell}}  d \cos \theta_{\ell} -  \int_{-1}^{0}  \frac{d \Gamma(\Lambda_b \to p \ell^- \bar{\nu}_{\ell})}{d q^2 d \cos \theta_{\ell}}  d \cos \theta_{\ell}}{\int_{0}^{1}  \frac{d \Gamma(\Lambda_b \to p \ell^- \bar{\nu}_{\ell})}{d q^2 d \cos \theta_{\ell}}  d \cos \theta_{\ell} + \int_{-1}^{0}  \frac{d \Gamma(\Lambda_b \to p \ell^- \bar{\nu}_{\ell})}{d q^2 d \cos \theta_{\ell}}  d \cos \theta_{\ell}} \,,\, \notag\\
    P_B(q^2) &=& \frac{d \Gamma^{\lambda_p=1/2} /{d q^2 } -d \Gamma^{\lambda_p=-1/2} /{d q^2 }  }{d \Gamma/{d q^2 }} ,\notag\\
    P_l(q^2) &=& \frac{d \Gamma^{\lambda_l=1/2} /{d q^2 } -d \Gamma^{\lambda_l=-1/2} /{d q^2 }  }{d \Gamma/{d q^2 }} .
\end{eqnarray}
The differential widths with definite polarization of the final state can be written as 
\begin{eqnarray}
	\frac{d \Gamma^{\lambda_p=1/2}}{d q^2 } &=&  \frac{4 m_{\ell}^2}{3q^2} \left (H^2_{\frac{1}{2},1} +H^2_{\frac{1}{2},0} +3H^2_{\frac{1}{2},t} \right ) + \frac{8}{3} \left (H^2_{\frac{1}{2},0} + H^2_{\frac{1}{2},1}\right ) \,,\,\notag\\
	\frac{d \Gamma^{\lambda_p=-1/2}}{d q^2 } &=&  \frac{4 m_{\ell}^2}{3q^2} \left (H^2_{-{1 \over 2},-1} +H^2_{-{1 \over 2},0} +3H^2_{\frac{-1}{2},t}\right ) + \frac{8}{3} \left (H^2_{\frac{-1}{2},0} + H^2_{\frac{-1}{2},-1}\right ) \,,\,\notag\\
    \frac{d \Gamma^{\lambda_l=1/2}}{d q^2 } &=&  \frac{ m_{\ell}^2}{q^2} \left[\frac{4}{3} \left (H^2_{\frac{1}{2},1} +H^2_{\frac{1}{2},0} +H^2_{\frac{1}{2},1} +H^2_{\frac{1}{2},0}\right ) + 4 \left ( H^2_{\frac{1}{2},t} +  H^2_{\frac{-1}{2},t}\right )  \right]\,,\,\notag\\
    \frac{d \Gamma^{\lambda_l=-1/2}}{d q^2 } &=&  \frac{8}{3} \left (H^2_{\frac{1}{2},1} +H^2_{\frac{1}{2},0} +H^2_{\frac{1}{2},1} +H^2_{\frac{1}{2},0}\right ) \, .
\end{eqnarray}

Using the form factors we obtained through combined fitting between our LCSR predictions and the lattice results, we can attain the following partially integrated decay rates of $ \Lambda_b \to p \mu^- \bar{\nu}_{\mu}$,
\begin{eqnarray}
	\zeta_{p \mu^- \bar{\nu}_{\mu}} (15 {\rm GeV^2})= \frac{1}{|V_{ub}|^2} \int_{15{ \rm GeV^2} }^{q^2_{max}}\frac{d \Gamma(\Lambda_b \to p \mu^- \bar{\nu}_{\mu})}{d q^2 } = \left( 11.80 \pm 2.56  \right) {\rm ps^{-1}}.
\end{eqnarray}   
We adopt the corresponding partially integrated decay rates of $ \Lambda_b \to \Lambda_c \mu^- \bar{\nu}_{\mu}$ from lattice QCD \cite{detmold_mathrmensuremathlambda_bensuremathrightarrowpensuremathellensuremath-overlineensuremathnu_ensuremathell_2015}, $\zeta_{\Lambda_c \mu^- \bar{\nu}_{\mu}} (7 {\rm GeV^2})$, and the experiment results from LHCb \cite{aaij_determination_2015} 
\begin{eqnarray}
	\zeta_{\Lambda_c \mu^- \bar{\nu}_{\mu}} ({\rm 7 GeV^2})= \left( 8.37 \pm 0.16 \pm 0.34  \right) {\rm ps^{-1}}\,,\, 
\end{eqnarray} 
\begin{eqnarray}
	 \frac{\int_{\rm 15 GeV^2 }^{q^2_{max}}\frac{d \Gamma(\Lambda_b \to p \mu^- \bar{\nu}_{\mu})}{d q^2 } }{\int_{7 {\rm GeV^2} }^{q^2_{max}}\frac{d \Gamma(\Lambda_b \to \Lambda_c \mu^- \bar{\nu}_{\mu})}{d q^2 } } =\left( 1.00 \pm 0.04 \pm 0.08 \right) \times 10^{-2}.
\end{eqnarray}
Taking the value of $|V_{cb}|= \left(39.5 \pm 0.8 \right) \times 10^{-3}$ extracted from exclusive $B$-decay \cite{aaij_determination_2015}, we predict the $|V_{ub}|$ as 
\begin{eqnarray}
	|V_{ub}|=\left(3.33\pm 0.43 \right) \times 10^{-3}.
\end{eqnarray}
Our extracted value of $|V_{ub}|$ remains approximately $1\sigma$ lower than the exclusive $B$-decay determination \cite{ParticleDataGroup:2024cfk}. 
However, if we adopt the inclusive determination $|V_{cb}| = (42.2 \pm 0.5) \times 10^{-3}$ from the same source \cite{ParticleDataGroup:2024cfk}, the resulting ratio yields $|V_{ub}| = (3.55 \pm 0.45) \times 10^{-3}$, which is notably closer to the exclusive $B$-decay value.
This comparison highlights that the uncertainty in $|V_{cb}|$, as well as the theoretical precision of the $\Lambda_b \to \Lambda_c \mu^- \bar{\nu}{\mu}$ decay-rate calculation, significantly affects the $|V_{ub}|$ extraction. Therefore, alongside further theoretical refinements of both $\Lambda_b \to p \mu^- \bar{\nu}{\mu}$ and $\Lambda_b \to \Lambda_c \mu^- \bar{\nu}{\mu}$, a complementary and promising approach is to determine $|V_{ub}|$ directly from future experimental measurements of the $\Lambda_b \to p \mu^- \bar{\nu}_{\mu}$ branching fraction.


Considering the life time of $\Lambda_b $, $\tau_{\Lambda_b}=1.470 {\rm ps}$  \cite{ParticleDataGroup:2024cfk}, and  $|V_{ub}|=\left(3.33\pm 0.43 \right) \times 10^{-3}$, we can numerically predict the total branching fractions,
the averaged forward-backward asymmetry $\left \langle  A_{\rm FB}  \right \rangle $ the averaged final hadron polarization $\left \langle  P_B  \right \rangle $ and the averaged lepton polarization $\left \langle  P_{\ell} \right \rangle $. 
The numerical results of the relevant observables in the semi-leptonic decays $\Lambda_b \to p \ell^- \bar{\nu}_{\ell}$ are presented in Table \ref{Pheno}. The predicted branching fraction for $\Lambda_b \to p \ell^- \bar{\nu}{\ell}$ is in reasonable agreement with results from lattice QCD \cite{PhysRevD.93.054003,detmold_mathrmensuremathlambda_bensuremathrightarrowpensuremathellensuremath-overlineensuremathnu_ensuremathell_2015} and from a heavy-hadron LCSR using the Ioffe current operator \cite{huang_lambda_2023}. However, the theoretical prediction of this observable depends intrinsically on the value of $|V_{ub}|$, which itself remains a significant source of uncertainty in the overall result.
We also list the averaged observables $\langle A_{\mathrm{FB}} \rangle$, $\langle P_B \rangle$, and $\langle P_{\ell} \rangle$, which are independent of $|V_{ub}|$.  The majority of our predictions are consistent with results obtained from other methods listed in \ref{Pheno}. 

\begin{table}[http]
\centering

\begin{tabular}{ccccccc}
\hline\hline
 &$\ell$& $\rm Br (\times 10^{-4}) $ & $\left \langle  A_{\rm FB}  \right \rangle$  & $\left \langle  P_B  \right \rangle$  & $\left \langle P_{\ell} \right \rangle$ \\
\hline
\multirow{3}{*}{This study} & $e$   &$4.1\pm 2.8$ &$0.34 \pm 0.13$&$-0.96\pm 0.19$ & $-1.000\pm 0.000$\\
 &$\mu$  &$4.1\pm 3.0$ &$0.34 \pm 0.12$&$-0.96\pm 0.19$ & $-0.995\pm 0.006$ \\
 &$\tau$ & $2.9\pm1.6$ & $0.42 \pm 0.07$ & $-0.95 \pm 0.11$ & $-0.54\pm 0.14$ \\ 
\hline\hline
\multirow{3}{*}{LCSR-Ioffe} \cite{huang_lambda_2023} & $e$ &$3.74 \pm 0.92$ &$0.33 \pm 0.01$&$-0.95\pm 0.05$ & $-1.00\pm 0.00$ \\
 &$\mu$   &$3.73 \pm 0.91$ &$0.32 \pm 0.01$&$-0.95\pm 0.05$ & $-0.99\pm 0.00$ \\
 &$\tau$ & $2.59\pm0.63$ & $0.15 \pm 0.001$ & $-0.93 \pm 0.04$ & $-0.55\pm 0.04$ \\ 
\hline
\multirow{3}{*}{RQM} \cite{PhysRevD.94.073008} & $e$ &4.5 &0.346&- & -0.91 \\
 &$\mu$   &4.5 &0.344&- & -0.91 \\
 &$\tau$ &2.9 & 0.185 & - & -0.89 \\ 
\hline
\multirow{2}{*}{LQCD} \cite{PhysRevD.93.054003} & $e(\mu)$ & 3.89&-&-&-  \\
&$\tau$ &2.74 &-&-&-\\
\hline
{light-LCSR} \cite{bell_light-cone_2013} & $e(\mu)$ &$4.0 ^{+2.3}_{-2.0} $&-&-&- \\
\hline\hline
\end{tabular}
\caption{Summary of the total branching fractions,
the averaged forward-backward asymmetry $\left \langle  A_{\rm FB}  \right \rangle $ the averaged final hadron polarization $\left \langle  P_B  \right \rangle $ and the averaged lepton polarization $\left \langle  P_l \right \rangle $, with the comparison with other works.}
\label{Pheno}\end{table}

\section{ Summary} \label{sect5}
We computed the radiative corrections to the $\Lambda_b \to p$ transition form factors at NLL accuracy within the framework of QCD light-cone sum rules, employing the distribution amplitudes of the $\Lambda_b$ baryon.
The factorization formulae were constructed by analyzing the form factors at leading power in $\lambda \sim m_p/m_{\Lambda_b}$ using the method of regions. The resulting structure demonstrated the validity of the factorization.  We adopted the LP current to construct the vacuum-to-$\Lambda_b$ correlation function. This choice simplified the calculation of the $\mathcal{O}(\alpha_s)$ perturbative corrections at leading power while still capturing the dominant behavior of the form factors.

Our analysis focused on the large-recoil region $q^2 < 8~\text{GeV}^2$, where the light-cone operator product expansion is applicable. The NLO hard kernel included only the hard and hard-collinear contributions. The use of the momentum-space light-cone projector for the $\Lambda_b$ baryon eliminated potential operator mixing. Furthermore, owing to the isospin invariance of the light-quark fields inside the $\Lambda_b$ baryon, all contributions involving the twist-3 distribution amplitudes canceled upon  summing over all diagrams.
 Consequently, the final expressions for the form factors depended solely on the twist-4 $\Lambda_b$ distribution amplitude.

 We  also justified that all 6 form factors  have the factorization-scale independence at one loop by computing convolution integrals of the NLO partonic distribution amplitude with
 the tree-level hard kernel .
  After  resummation of large logarithms, we attained $\Lambda_b \to p $ form factors at NLL accuracy.
	  
  For our numerical analysis, we primarily adopted the exponential model $\psi_4^{\mathrm{I}}$ to parameterize the $\Lambda_b$-baryon distribution amplitude $\psi_4(\omega, \mu)$ at a soft scale. 
  To assess model dependence, we also employed two alternative parameterizations, $\psi_4^{\mathrm{II}}(\omega,\mu)$ and $\psi_4^{\mathrm{III}}(\omega,\mu)$.
The numerical results indicated that the resummation of parametrically large logarithms in the hard matching coefficient has a relatively minor effect on the form factors compared to the pure one-loop fixed-order correction. 
The perturbative $\mathcal{O}(\alpha_s)$ correction is dominated by the contributions from the NLO jet function. This correction reduces the tree-level sum-rule prediction  to approximately 65$\%$ of their original value , highlighting the significant role of QCD radiative effects in baryonic sum-rule applications. 
Furthermore, both the LL and NLL sum-rule predictions clearly exhibit the expected $1/E_p^3$ scaling at large recoil, consistent with the power-counting analysis.
In addition, a first-order expansion in the $z$-parameter provides an excellent fit to our NLL results, mirroring the behavior observed in the $\Lambda_b \to \Lambda$ process \cite{wang_perturbative_2016} and in numerous $B$-meson decays \cite{wang_subleading_2017, wang_qcd_2015}. 
These common features may point to a shared underlying mathematical structure for these heavy-to-light form factors.
After converting our results to the standard $V-A$ basis, a comparison with other calculations showed that while our leading-power analysis with the LP current captures the dominant behavior of the form factors, higher-power corrections still exert a substantial influence.

We also performed  a correlated $\chi^2$ fit to combine our form-factor predictions with the lattice QCD results from \cite{detmold_mathrmensuremathlambda_bensuremathrightarrowpensuremathellensuremath-overlineensuremathnu_ensuremathell_2015}. 
Using the fitted form factors, we calculated the partially integrated decay rates for $\Lambda_b \to p \mu^- \bar{\nu}_{\mu}$ and extracted
 $|V_{ub}|=\left(3.33\pm 0.43 \right) \times 10^{-3}.$ 
It should be noted that the extracted value depends on the input for $|V_{cb}|$ and on the theoretical precision of the $\Lambda_b \to \Lambda_c \mu^- \bar{\nu}{\mu}$ decay rate, which together constitute the dominant systematic uncertainty. A more direct and robust determination of $|V_{ub}|$ would follow from a future experimental measurement of the $\Lambda_b \to p \mu^- \bar{\nu}_{\mu}$ branching fraction, thereby circumventing these external dependencies.
 In addition, we further investigated the  phenomenological applications of $\Lambda_b \to p \ell^- \bar{\nu}_{\ell}$ process, such as the  predictions for the total branching fraction, the averaged forward–backward asymmetry $\langle A_{\mathrm{FB}} \rangle$, the averaged hadron polarization $\langle P_B \rangle$, and the averaged lepton polarization $\langle P_{\ell} \rangle$. 

While the evaluated NLL QCD correction markedly enhances the accuracy of the form-factor predictions, the neglected power-suppressed contributions from the heavy-quark expansion still give rise to significant systematic uncertainties. To achieve further improvements in precision, higher-twist $\Lambda_b$ distribution amplitudes and potential non-factorizable corrections  must also be taken into account.
In addition, the improved model  of the twist-4 $\Lambda_b$-baryon LCDA with the radiative tail at large $\omega$ due to perturbative corrections \cite{feldmann_radiative_2025}  may have implications.
A systematic investigation of these contributions will be pursued in future work.

\section*{Acknowledgements}
This work acknowledges support from the National Natural Science
Foundation of China with Grants No. 12475097 and No.12535006, and from the Natural Science Foundation of
Tianjin with Grant No. 25JCZDJC01190.

\appendix

\bibliographystyle{JHEP}
\bibliography{reference}
\end{document}